# Giant phonon softening and avoided crossing in aliovalence-doped heavy-band thermoelectrics

Shen Han[1,†], Shengnan Dai[2,†], Jie Ma[3,†], Qingyong Ren[4,5], Chaoliang Hu[1], Ziheng Gao[1], Manh Duc Le[6], Denis Sheptyakov[7], Ping Miao[4,5,8], Shuki Torii[8], Takashi Kamiyama[8], Claudia Felser[9], Jiong Yang[2,10*], Chenguang Fu[1*], Tiejun Zhu[1*]

[1]State Key Laboratory of Silicon Materials, School of Materials Science and Engineering, Zhejiang University, 310027 Hangzhou, China.

[2]Materials Genome Institute, Shanghai University, 99 Shangda Road, 200444 Shanghai, China.

[3]Key Laboratory of Artificial Structures and Quantum Control, School of Physics and Astronomy, Shanghai Jiao Tong University, 800 Dongchuan Road, 200240 Shanghai, China.

[4]Institute of High Energy Physics, Chinese Academy of Sciences, 100049 Beijing, China.

[5]Spallation Neutron Source Science Center, 523803 Dongguan, China.

[6]ISIS Neutron and Muon Source, STFC Rutherford Appleton Laboratory, Chilton, Didcot, OX11 0QX Oxfordshire, United Kingdom

[7]Laboratory for Neutron Scattering and Imaging, Paul Scherrer Institut, 5232 Villigen, Switzerland

[8]Institute of Materials Structure Science, High Energy Accelerator Research Organization (KEK), Tokai, Ibaraki 319-1106, Japan

[9]Max Planck Institute for Chemical Physics of Solids, Nöthnitzer Straße 40, Dresden 01187, Germany.

[10]Zhejiang Laboratory, Hangzhou, Zhejiang, 311100, China.

[†]These authors contributed equally: Shen Han, Shengnan Dai, Jie Ma.

[*]email: chenguang_fu@zju.edu.cn; jiongy@t.shu.edu.cn; zhutj@zju.edu.cn

**Abstract**

　　Aliovalent doping has been adopted to optimize the electrical properties of semiconductors, while its impact on the phonon structure and propagation is seldom paid proper attention to. This work reveals that aliovalent doping can be much more effective in reducing the lattice thermal conductivity of thermoelectric semiconductors than the commonly employed isoelectronic alloying strategy. As demonstrated in the heavy-band NbFeSb system, a large reduction of 65% in the lattice thermal conductivity is achieved through only 10% aliovalent Hf-doping, compared to the 4 times higher isoelectronic Ta-alloying. It is elucidated that aliovalent doping introduces free charge carriers and enhances the screening, leading to the giant softening and deceleration of optical phonons. Moreover, the heavy dopant can induce the avoided-crossing of acoustic and optical phonon branches, further decelerating the acoustic phonons. These results highlight the significant role of aliovalent dopants in regulating the phonon structure and suppressing the phonon propagation of semiconductors.

**Introduction**

Thermoelectric (TE) materials, which can realize the direct conversion between heat and electricity, exhibit promising applications in waste heat recovery and refrigeration applications[1], providing a way for increasing energy use efficiency and achieving carbon neutrality[2]. The energy conversion efficiency for TE materials is quantified by the dimensionless figure of merit, $zT=S^2\sigma T/(\kappa_e+\kappa_L)$, where $S$, $\sigma$, $T$, $\kappa_e$, and $\kappa_L$ are the Seebeck coefficient, the electrical conductivity, the absolute temperature, the electronic and lattice components of thermal conductivity, respectively[3]. The challenge in the optimization of TE performance lies in the intercoupling of the above parameters. Among them, the carrier concentration $n$ plays a vital role in modulating the electrical transport parameters.

For heavily doped TE semiconductors, the increase of $n$ means that the Fermi level moves inside either the conduction band (CB) or valence band (VB), generally contributing to the higher $\sigma$ and $\kappa_e$ because of the increased electronic density of states (DOS), whereas, the differential conductivity near the Fermi level will become less asymmetric, reducing $S$[4]. Additionally, the carrier mobility $\mu$, determined by the dominant scattering mechanism, also exhibits a significant $n$-dependence[5,6]. Owing to the different $n$-dependences in $\sigma(n)$, $\mu(n)$, and $S(n)$, the $zT$ reaches the maximum only at the optimal carrier concentration $n_{opt}$. Under the classical statistics approximation, the $n_{opt}$ is approximately proportional to $(m_d^* T)^{3/2}$ in a single-band system[7], where $m_d^*$ ($m_d^* = N_v^{3/2} m_b^*$) is the density of states effective mass, $N_v$ the band degeneracy and $m_b^*$ the band effective mass (see Fig. S1 for a schematic illustration of optimal carrier concentrations at different DOS effective masses). For various TE materials, the large difference in $m_d^*$ can result in significantly distinct $n_{opt}$ (Fig. 1a), requiring different doping strategies. The conventional TE materials, i.e. $Bi_2Te_3$, Pb(Te, Se), exhibit s-orbital and/or p-orbital dominated CB and VB, resulting in a small band effective mass $m_b^*$ ($< 1m_e$) and $n_{opt}$ ($\sim 10^{19}$ cm$^{-3}$). In contrast, more recently developed TE materials, like half-Heusler (HH) compounds and skutterudites, show a much larger $m_b^*$ ($> 1m_e$) owing to the d-orbital contributed CB and VB, requiring higher doping content for achieving $n_{opt}$ ($10^{20}$-$10^{21}$ cm$^{-3}$) and the optimal $zT$ (Fig. 1a).

HH compounds with 18 valence electrons have attracted extensive attention in recent years as promising TE materials for power generation[8,9], exhibiting high $zT$ above unity in both n-type[10,11] and p-type[12,13]. Recently developed TE modules using n-type (Zr, Hf)NiSn and p-type NbFeSb-based HH compounds achieve a high conversion efficiency of 10.5% together with a large power density of 3.1 W cm$^{-2}$ at a temperature difference of 680 K[14], demonstrating the prospect of HH compounds for power generations. Distinct from the conventional light-band ($m_b^* < 1m_e$) TE materials, heavy-band ($m_b^* > 1m_e$) HH compounds intrinsically possess both large $m_d^*$ and $\kappa_L$. The former results in the high $n_{opt}$, e.g., $\sim 4 \times 10^{20}$ cm$^{-3}$ for n-type ZrNiSn[15] and $\sim 2.6 \times 10^{21}$ cm$^{-3}$ for p-type NbFeSb[16], which are one or two orders of magnitude higher than that of PbTe ($\sim 3 \times 10^{19}$ cm$^{-3}$)[17], respectively. High $n_{opt}$ indicates the demand for high content of aliovalent doping to optimize the electrical properties. For a HH compound with a face-centered cubic structure and lattice constant of 6 Å, a nominal doping concentration of 10% is required for realizing the $n_{opt}$ of $2 \times 10^{21}$ cm$^{-3}$, if assuming that each dopant provides one electron (i.e., a doping efficiency of 100%). In practice, the doping efficiency is generally lower than 100%, suggesting that an even higher doping content is required to achieve the optimal $zT$. Such a high doping content ($\sim 10\%$) will unavoidably affect the $\kappa_L$ of the matrix.

Fig. 1b shows the effect of aliovalent doping on the $\kappa_L$ of both light-band and heavy-band systems. Although leading to the suppression of $\kappa_L$ in both cases, it is much more significant in heavy-band systems owing to the higher amount of doping needed to reach $n_{opt}$. For instance, a large reduction of 65% in the $\kappa_L$ of heavy-band NbFeSb is obtained with 10% Hf-doping. In contrast, isoelectronic alloying is a more commonly used strategy to reduce the $\kappa_L$ by introducing point defect scattering[18,19], especially for the heavy-band HH compounds[20–22]. The effect of aliovalent doping and isoelectronic alloying on the $\kappa_L$ of TE materials is presented in Fig. 1b. It is unexpectedly found that aliovalent doping is much more effective in the suppression of $\kappa_L$, compared to isoelectronic alloying. For instance, 10% aliovalent doping is enough to realize a 50% reduction in the $\kappa_L$, whereas 30% isoelectronic alloying is needed for the same $\kappa_L$ reduction. To effectively suppress the $\kappa_L$, aliovalent doping is thus much preferable to isoelectronic alloying, besides the synergistic optimization of electrical properties. Bearing this in mind, it is intriguing to uncover why aliovalent doping is so effective in reducing the $\kappa_L$.

Here, taking the heavy-band NbFeSb as an example, the mechanisms that aliovalent doping and isoelectronic alloying suppress the phonon transport were investigated using various probes, including electrical and thermal transport measurements, inelastic neutron scattering (INS) measurements, and first-principles calculations. We reveal that the aliovalent dopants induce a distinct softening of optical phonons and also decelerate the optical phonons. Moreover, avoided crossing of optical and acoustic phonon branches is found when doping with heavy elements, which further decelerates the acoustic phonons. These results provide new insights into how aliovalent doping could significantly modulate the phonon dispersion of TE materials.

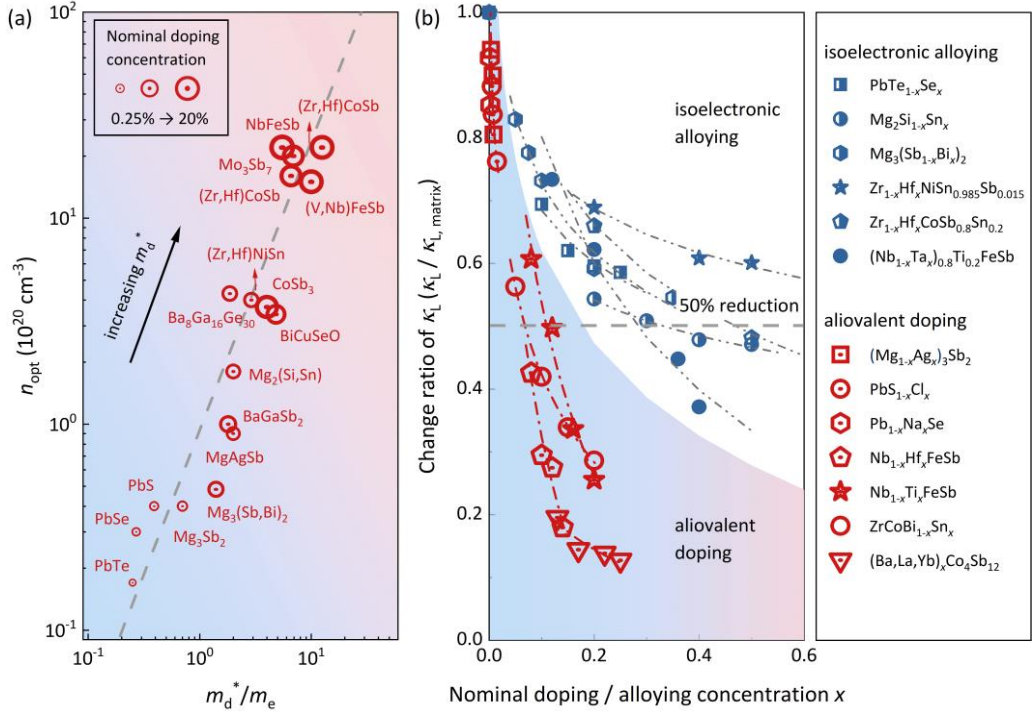

**Fig. 1. The optimal carrier concentration and suppression of lattice thermal conductivity via aliovalent doping for typical TE materials. a** The optimal carrier concentration versus the density of states effective mass, the size of the dot indicates the nominal doping concentrations in experiments. **b** The reduction of lattice thermal conductivity versus the doping / alloying concentration. (See Table S1-S3 in the Supplementary Information for detailed data and references.)

## Results

**Thermal transport properties.** High-quality $Nb_{0.9}X_{0.1}FeSb$ ($X$ = Nb, Ta, Hf, Zr) polycrystalline samples were synthesized by levitation melting and spark plasma sintering (see Methods for details and Table S4 for the composition and relative density for all samples). No obvious impurity phase is found for all the polycrystalline samples from the neutron powder diffraction patterns (Fig. S2). The temperature dependences of the total thermal conductivity $\kappa$, $S$, $\sigma$ were measured over the temperature range of 2 K to 1000 K (Fig. S3). The $\kappa_L$ was calculated by subtracting the electronic component $\kappa_e$, which was estimated via the Wiedemann-Franz relationship $\kappa_e = L\sigma T$, where $L$ is the Lorenz number calculated by the single parabolic band (SPB) approximation[23]. As shown in Fig. 2a, the $\kappa_L$ of all aliovalent Hf-/Zr-doped and isoelectronic Ta-alloyed samples shows an obvious reduction, while the maximum drop is found in the sample with 10% Hf-doping, exhibiting a large reduction of ~70% at 100 K and ~65% at 300

K. This large drop in the $\kappa_L$ at a doping concentration 10% is highly unexpected, considering the effectiveness of isoelectronic alloying in reducing the $\kappa_L$, for instance, an 50% isoelectronic Hf-alloying only leads to a 40% reduction in the $\kappa_L$ of (Zr,Hf)NiSn[21].

First-principles calculations were carried out to unravel the mechanism of how doping and alloying affect the phonon-phonon scattering in the heavy-band NbFeSb system. Here a doping/alloying concentration of 12.5% was selected to facilitate the calculations. The phonon DOS, phonon dispersions, and the $\kappa_L$ were obtained by the Phonopy package[24] and shengBTE package[25] (see Methods for details). First, the $\kappa_L$ of Nb$_{0.875}$X$_{0.125}$FeSb was calculated by only considering the phonon-phonon scattering (Fig. 2b). For NbFeSb, the calculated $\kappa_L$ at 300 K is 23.9 W m$^{-1}$ K$^{-1}$, which is in reasonable agreement with the experimental value (17.1 W m$^{-1}$ K$^{-1}$) obtained in the polycrystalline sample (Fig. 2a). With Hf-doping, there is a 40% drop in $\kappa_L$ for Nb$_{0.875}$Hf$_{0.125}$FeSb, suggesting Hf-doping can significantly enhance the phonon-phonon scattering. Since the aliovalent Hf-doping will also induce a large increase of $n$, we further introduce electron-phonon scattering to the calculations using the Quantum ESPRESSO[26] and EPW[27] packages, the $\kappa_L$ at 300 K reduces to 12.6 W m$^{-1}$ K$^{-1}$, a reduction of ~47% compared to that of NbFeSb (Table S5). Such a reduction approaches the observed drop in the experimental polycrystalline samples, suggesting that first-principles calculations could provide insights into how the aliovalent doping affects phonon transport.

Fig. 2c shows the calculated phonon energy-dependent $\kappa_L$ for Nb$_{0.875}$X$_{0.125}$FeSb at 300 K through the expression as[28]

$$\kappa_L = \frac{1}{3}\sum C_v v_g^2 \tau \qquad (1)$$

where $C_v$ is the heat capacity, $v_g$ the group velocity, and $\tau$ the relaxation time. One finds that the drop of $\kappa_L$ mainly happens in the energy range of 10 meV to 30 meV when only considering the phonon-phonon scattering. Moreover, by further considering the electron-phonon scattering, an additional drop of $\kappa_L$ in the energy range below 10 meV is observed. Generally, the drop in the $\kappa_L$ of a solid originates from either the reduced

$v_g$ or $\tau$, or both. To further distinguish which is the dominant factor leading to the large reduction in $\kappa_L$ of $Nb_{0.875}X_{0.125}FeSb$, we recalculate the $\kappa_L$ using eq. (1) by considering both changes in $v_g$ and $\tau$ (i), only changing $\tau$ (ii), and only changing $v_g$ (iii), as shown in Fig. 2d. Considering only the change of $\tau$ (ii), the isoelectronic Ta alloying leads to a slight drop of $\kappa_L$ (4%) while the aliovalent doping of Hf and Zr results in a 10% drop of $\kappa_L$. However, when compared with case (i), only considering the change of $\tau$ (ii) cannot reproduce the large drop of $\kappa_L$. Meanwhile, the calculated results by only changing $v_g$ are presented (iii), which can well reproduce the large drop of $\kappa_L$, suggesting that both aliovalent doping and isoelectronic alloying in $Nb_{0.875}X_{0.125}FeSb$ play a vital role in decelerating the group velocity rather than the relaxation time. Hence, there should be a significant change in the phonon dispersion and density of states (DOS) of $Nb_{0.875}X_{0.125}FeSb$ compared with NbFeSb.

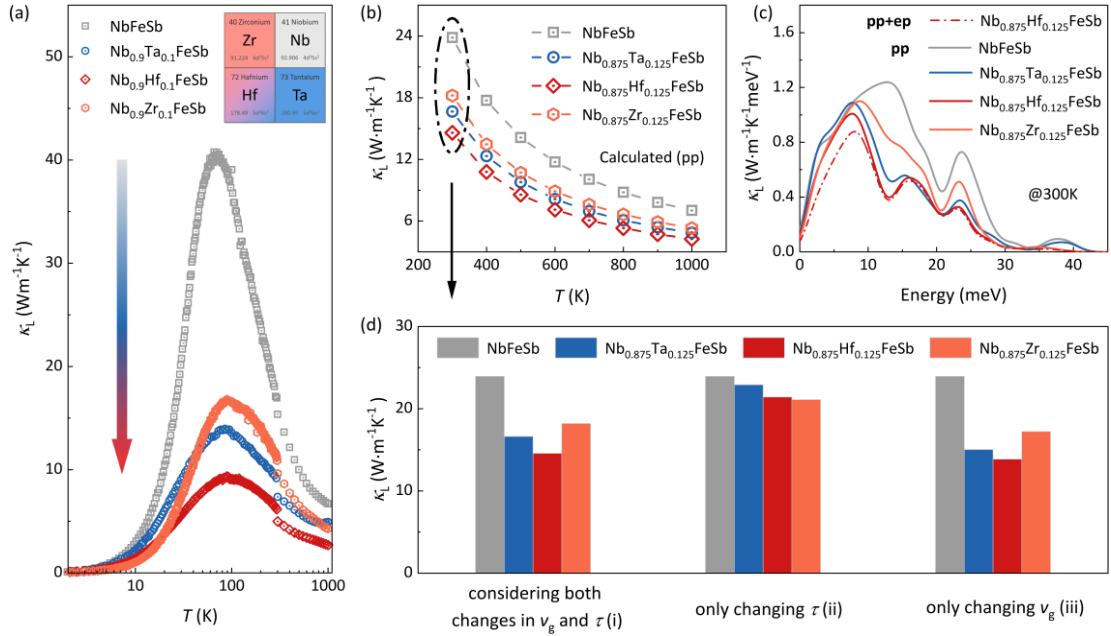

**Fig. 2. Comparison of the $\kappa_L$ of NbFeSb with either aliovalent doping or isoelectronic alloying. a** Temperature-dependent $\kappa_L$ for $Nb_{0.9}X_{0.1}FeSb$ polycrystalline samples. **b** The calculated $\kappa_L$ for $Nb_{0.875}X_{0.125}FeSb$ by only considering phonon-phonon scattering. **c** Phonon energy-dependent $\kappa_L$ for $Nb_{0.875}X_{0.125}FeSb$ at 300 K. Only phonon-phonon scattering (pp) was considered for the solid lines, while the dashed line is the data with further consideration of electron-phonon scattering (ep). Detailed data for $Nb_{0.875}Hf_{0.125}FeSb$ and $Nb_{0.875}Zr_{0.125}FeSb$ with ep are shown in Fig. S4 and Table S5. **d** The calculated $\kappa_L$s for $Nb_{0.875}X_{0.125}FeSb$ at 300K using eq. (1) by considering both changes in $v_g$ and $\tau$ (i), only changing $\tau$ (ii), and only changing $v_g$ (iii).

**Aliovalent doping-induced optical phonon softening.** To understand the roles of doping and alloying in the changes of phonon dispersion, phonon density of states (DOSs) measurements for $Nb_{0.9}X_{0.1}FeSb$ using the INS technique were carried out on the MARI time-of-flight chopper spectrometer at the ISIS Neutron and Muon Source in the UK. The momentum and energy dependence of the powder averaged dynamical structure factor $S(Q, E)$ was recorded. Fig. 3a shows the results for $Nb_{0.9}X_{0.1}FeSb$ polycrystalline samples at 5 K as examples (See Fig. S5 for results at different temperatures). The neutron-weighted phonon DOSs were obtained by integration of the $S(Q, E)$ over the whole momentum transfer range (see methods). Fig. 3b shows the neutron-weighted phonon DOSs of $Nb_{0.9}X_{0.1}FeSb$ at 5 K (See Fig. S6 for results at different temperatures). Three phonon bands, i.e., 0-20 meV, 20-30 meV, and 30-45 meV, are observed, corresponding to one acoustic phonon band and two optical phonon bands, respectively. A marked softening of the optical phonons at the energy range of 30-35 meV is observed for the Hf- and Zr-doped $Nb_{0.9}Hf_{0.1}FeSb$ and $Nb_{0.9}Zr_{0.1}FeSb$, which however does not happen for the isoelectronic Ta-alloyed $Nb_{0.9}Ta_{0.1}FeSb$ (Fig. 3b). To verify the observed shift, the calculated phonon DOSs of $Nb_{0.875}X_{0.125}FeSb$ is also presented. The calculations match well with the experimental data by exhibiting a marked softening of high-energy optical phonons, indicating that the aliovalent doping of Hf and Zr in NbFeSb brings a distinct effect on the phonon structure compared to the isoelectronic alloying of Ta.

The change in phonon energy will affect the distribution of phonon modes, which in turn modulates the group velocities and scattering process. Fig. 3c shows the calculated group velocities of phonon modes in the energy range of 20-45 meV (corresponding to optical phonons) in $Nb_{0.875}X_{0.125}FeSb$ compared with that in the pure NbFeSb. Significant suppression of group velocities in optical phonons for aliovalent Hf- and Zr-doped compositions appeared in the energy range 20-40 meV, corresponding to the softening of optical phonons, but this does not occur for the isoelectronic Ta-alloyed $Nb_{0.875}Ta_{0.125}FeSb$, suggesting the distinct impact of aliovalent doping. The softening of the optic modes suppresses the group velocity ($v_g = d\omega/dk$) through the combination of two effects: On one hand, the optic branches are flattened (Fig. S7), while on the other hand, the phonon softening also pushes a portion of high-energy phonon modes to lower energies. The suppression of group velocity due to aliovalent doping acts on the heat-transferred optical phonons, contributing to the suppressed $\kappa_L$.

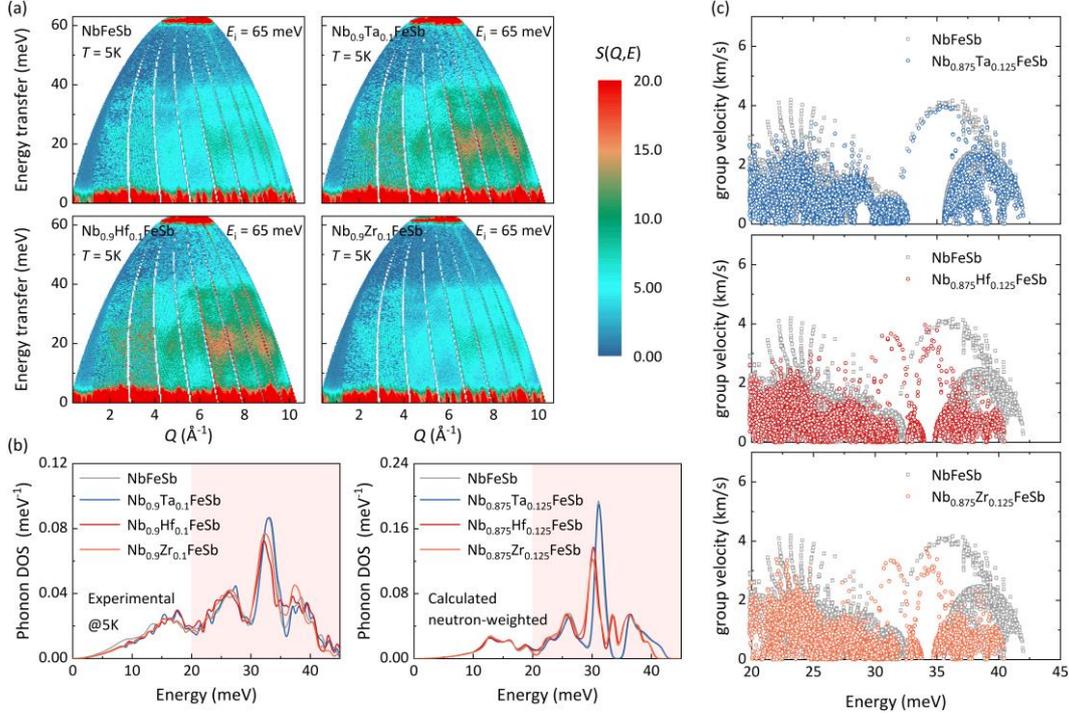

**Fig. 3. The softening of optical phonons and suppressed group velocity. a** The experimental dynamical structure factor, $S(Q, E)$, measured using the INS technique for $Nb_{0.9}X_{0.1}FeSb$ polycrystalline samples at 5 K with $E_i$ = 65 meV. **b** The measured neutron-weighted phonon DOSs of the $Nb_{0.9}X_{0.1}FeSb$ compounds and the calculated neutron-weighted phonon DOSs of $Nb_{0.875}X_{0.125}FeSb$. **c** Energy dependences of calculated group velocity for $Nb_{0.875}X_{0.125}FeSb$ compared with NbFeSb (black hollow) in the energy range of 20-45 meV.

The positions in the periodic table and valence electron information of the matrix element Nb, the alloying element Ta, and the aliovalent doping elements Zr and Hf are shown in the inset of Fig. 2a. Compared with Ta in the VB group, Hf and Zr, as IVB-elements, introduce free charge carriers to the NbFeSb matrix. Fig. 4a presents the calculated phonon dispersion in NbFeSb with the total and partial phonon DOS. The degeneracy of longitudinal optical (LO) phonon and transverse optical (TO) phonon at ~32 meV is lifted around the Brillouin zone center, corresponding to the LO-TO splitting, which matches with the decomposed two peaks of the 30-45 meV optical phonon band in both the experimental and the calculated phonon DOSs. The LO-TO splitting is caused by the repulsive electric (Coulomb) interaction between ions in a polar material, which tends to push them apart[29]. This raises the energy of the LO optic mode where the ions alternately move towards and away from each other. In previous work on the phonon dispersion of HH compounds, the LO-TO splitting in ZrNiSn has been reported[5] and this confirms the existence of polar chemical bonds in HH compounds[30], associated with the optical phonon scattering in electrical transport. Additional Coulomb interaction acts as the restoring force of the LO vibrations. Sb doping in ZrNiSn brings free carriers, which screen the Coulomb interaction and result in the collapse of the LO-TO splitting[5]. We now turn to the details of chemical bonding in NbFeSb to give a deeper understanding of the aliovalent-doping-induced phonon softening.

Compared with NbFeSb and $Nb_{0.875}Ta_{0.125}FeSb$, the high-energy optical phonon band (>30 meV) drops and the LO-TO splitting collapses in $Nb_{0.875}Hf_{0.125}FeSb$ and $Nb_{0.875}Zr_{0.125}FeSb$ (Fig. 4b). Partial phonon DOS of NbFeSb is presented in Fig. 4a, which shows that the high-energy optical phonon band is mainly contributed by Fe. It is worth considering the changes in the Fe-Nb bonds after the aliovalent doping with Hf and Zr in Nb site. The crystal orbital Hamiltonian population (COHP) was then calculated with the LOBSTER software[31–34] (Fig. S8). The negative integrated partial COHP (-IpCOHP) of the nearest neighbor (Nb, Hf, Zr)-Fe bonds in $Nb_{0.875}X_{0.125}FeSb$ is shown in Fig. 4c. The -IpCOHP probes the relative bond strength and a larger value indicates a stronger bonding. As seen in Fig. 4c, the -IpCOHP values for the nearest neighbor Nb-Fe bonds in Hf and Zr doped compositions are smaller than that in NbFeSb, suggesting that the introduction of free carriers enhances the screening and weakens the Nb-Fe bonds, resulting in the softening of optical phonons. It is also found that the -IpCOHP values of both the nearest neighbor Hf-Fe bond and Zr-Fe bond are smaller than that of the Nb-Fe bond, probably due to the difference in the initial total number of electrons.

Fig. 4d shows the calculated deformation charge density, obtained by subtracting the charge density of NbFeSb from that of $Nb_{0.875}Hf_{0.125}FeSb$ composition, to clarify the differences in the electron transfer in bonding. For the Nb-Fe bond, the positive values near the Nb atom indicate that the degree of electron transfer becomes weaker after Hf doping. For the Hf-Fe bond, although the Hf atom has one less electron than Nb, positive values of the deformation charge density are observed near the Fe atom. The decreased electron transfer during the bonding process means the weakening of bonds, corresponding to the above observed softening of optical phonons.

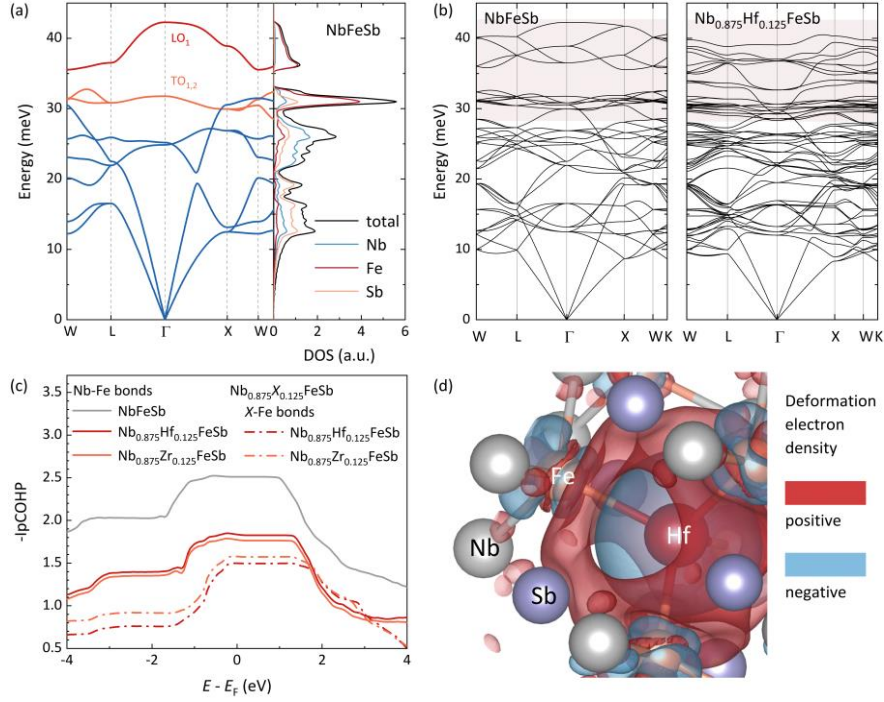

**Fig. 4. Weakened chemical bonding after aliovalent doping in NbFeSb.** Calculated phonon dispersion for **a** NbFeSb with the total and partial phonon DOS, **b** NbFeSb and $Nb_{0.875}Hf_{0.125}FeSb$ in a 2 × 2 × 2 supercell. **c** The negative integrated pCOHP (-IpCOHP) for the nearest neighbor (Nb, Hf, Zr)-Fe bonds in $Nb_{0.875}X_{0.125}FeSb$. **d** Deformation electron density in $Nb_{0.875}Hf_{0.125}FeSb$ compared with NbFeSb at an isosurface value of 0.0004 $e/bohr^3$. Lattice relaxation after doping is ignored because the exact corresponding lattice position is required for the deformation process. Red and blue correspond to positive and negative values, respectively.

**Heavy elements-induced avoided crossing.** The enhanced screening from the introduced free charge carriers in $Nb_{0.9}Hf_{0.1}FeSb$ and $Nb_{0.9}Zr_{0.1}FeSb$ leads to the softening and deceleration of optical phonons, suppressing the $\kappa_L$. It is worth noting that the isoelectronic Ta-alloyed $Nb_{0.9}Ta_{0.1}FeSb$ does not show a significant softening of the optical phonons but exhibits a lower $\kappa_L$ than that of $Nb_{0.9}Zr_{0.1}FeSb$. Moreover, a similar softening of optical phonons in the Hf- and Zr-doped NbFeSb can be found according to the phonon DOS and the IpCOHP, but the $\kappa_L$ of Hf-doped NbFeSb is much lower. Since the period-6 elements, Hf and Ta, have larger atomic masses than the

period-5 elements, Nb and Zr, it is thus interesting to consider how the heavy-element substitution affects the phonon transport in heavy-band TE semiconductors.

Previously, the effect of alloying on $\kappa_L$ in TE materials was generally considered as point defect scattering and analyzed using the Debye-Callaway model[35], while the quantitative effect is only attributed to the change of relaxation time $\tau$ under the perturbation theory without considering the change of phonon structure and related manifestations, such as group velocity or sound velocity. Aliovalent doping is here shown to suppress the group velocity of high-energy optical phonons, and isoelectronic alloying of elements with higher atomic mass may modulate low-energy acoustic phonons. Fig. 5a showed the calculated group velocities of phonon modes in the energy range of 0-25 meV in $Nb_{0.875}X_{0.125}FeSb$ and the results for different doping and alloying elements were compared with that in the pure NbFeSb. A significant drop in the group velocity of Ta-alloyed and Hf-doped $Nb_{0.875}X_{0.125}FeSb$ is observed in the energy range of 7-15 meV while it is not obvious in Zr-doped $Nb_{0.875}X_{0.125}FeSb$.

For acoustic phonons, the significant drop in the group velocity is directly related to the energy changes of acoustic branches. Therefore, we re-examine the calculated phonon dispersion for $Nb_{0.875}X_{0.125}FeSb$. Fig. 5b shows the phonon dispersion of NbFeSb, $Nb_{0.875}Zr_{0.125}FeSb$, and $Nb_{0.875}Hf_{0.125}FeSb$ below 25 meV. The acoustic branches change less in $Nb_{0.875}Zr_{0.125}FeSb$ and cross normally with optical branches, while a significant drop of acoustic branches and avoided-crossing behaviors between the acoustic and optical branches was observed in $Nb_{0.875}Hf_{0.125}FeSb$. A similar phenomenon occurs also in $Nb_{0.875}Ta_{0.125}FeSb$, as shown in Fig S5.

The avoided-crossing behavior was originally reported in structures with open-cage frameworks with guest fillers, e.g., Clathrates and filled skutterudites, explained as the independent rattling motion of guest fillers[36,37]. The guest fillers selected in the filled skutterudites are usually heavy elements, like Cs, Ba, and lanthanides, due to the lower energy of the rattling modes. The calculated partial phonon DOS for $X$ in $Nb_{0.875}X_{0.125}FeSb$ in the energy range of 0-25 meV is shown in Fig. 5c (See Fig. S9 and S10 for different elements results in the full energy range). Ta and Hf show much larger contributions to the DOS than Zr in this energy range, which corresponds well to the positions where the phonon group velocity drops. The experimental phonon DOSs of $Nb_{0.9}Ta_{0.1}FeSb$ and $Nb_{0.9}Hf_{0.1}FeSb$ also support this analysis by showing enhancements at around 20 meV (Fig. 3b).

The mass difference between the matrix and the heavy doping/alloying elements drives the occurrence of avoided crossing and the separation degree between acoustic and optical branches also increases with the increased mass of the doping/alloying atom, as theoretically analyzed (see details in the Supplementary Information) using the linear chain model[38]. Such behavior of doping/alloying heavy elements significantly reduces the group velocity of acoustic phonons, which in turn suppresses $\kappa_L$. To our knowledge, such an avoided crossing of phonon branches induced by heavy doping/alloying elements was previously not observed in TE systems. Considering that isoelectronic alloying with heavier elements is very commonly used to suppress the $\kappa_L$ of TE materials, the observation in this work could provide insights into the understanding of the suppressed phonon transport in other high-performance TE solid solutions.

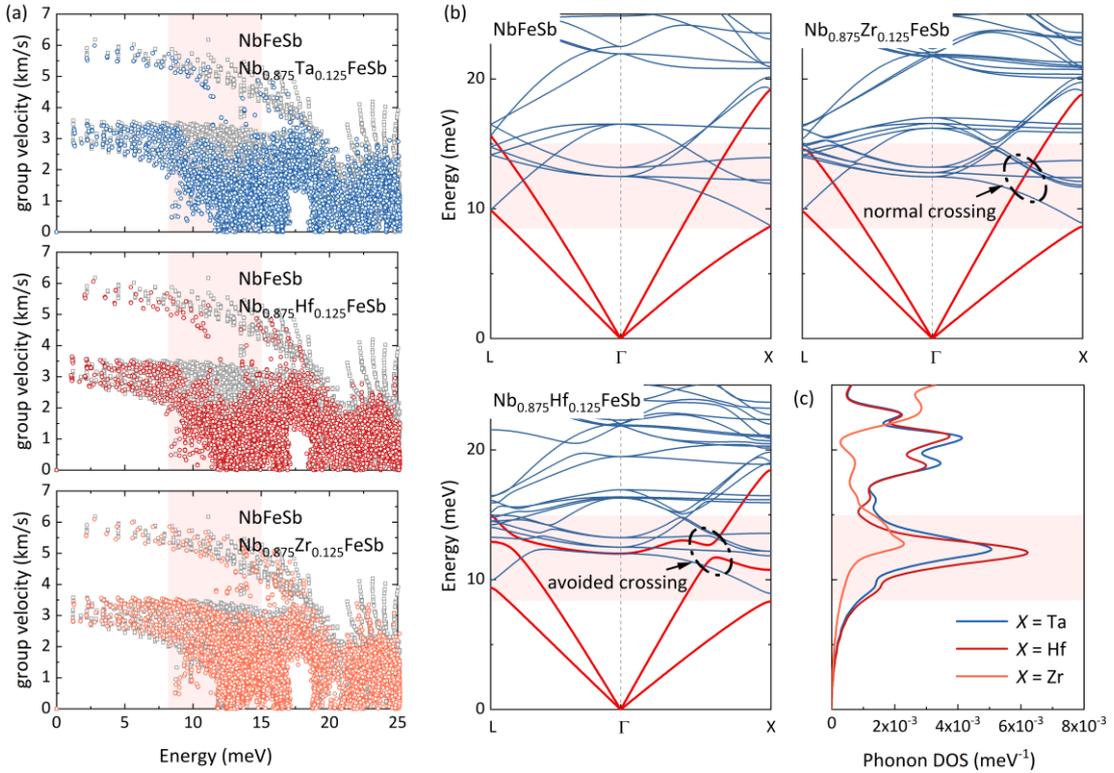

**Fig. 5. Avoided crossing behavior and the suppression of sound velocity. a** Energy dependences of calculated group velocity for $Nb_{0.875}X_{0.125}FeSb$ compared with NbFeSb (black hollow); **b** calculated phonon dispersion for NbFeSb, $Nb_{0.875}Zr_{0.125}FeSb$ and $Nb_{0.875}Hf_{0.125}FeSb$; **c** partial phonon DOS for $X$ in $Nb_{0.875}X_{0.125}FeSb$ in the energy range of 0-25 meV.

**Discussion**

For the analysis of the $\kappa_L$ of crystalline solids, the classical Debye-Callaway model is derived using perturbation theory with an assumption that the phonon dispersion in the crystal is unchanged[39–41], and the additional extrinsic scattering mechanisms, such as point defects, grain boundaries, dislocations, etc, are considered to only affect the phonon relaxation time[35]. The expression of relaxation time for intrinsic three-phonon scattering has been given individually with the differences in frequency and temperature dependency by Slack[42]. For the point defect scattering, the mass fluctuation and strain field fluctuation due to the mass and atomic size differences between impurity and the matrix atoms were considered to be the source of thermal resistance, and the expression of phonon relaxation time versus frequency and temperature has been given by Slack and Abeles[43–45]. The Debye-Callaway model, which could provide reasonable interpretations and guidelines for experimental studies is generally acceptable. However, in real cases, doping and alloying elements, particularly when reaching a considerable content, might significantly modulate the phonon dispersion and thus the group velocity.

Relying on the improvement of calculation resources and the development of the INS technique, it is now possible to reveal the doping- or alloying-induced change in the phonon dispersion and intrinsic three-phonon scattering of crystalline solids. Optical phonons with higher energies and lower group velocities were underappreciated previously, but fluctuations in their group velocities can affect thermal transport efficiently due to their large phonon DOSs. The group velocity of optical phonons is sensitive to the energy variation of optical phonon branches, meaning the softening of optical phonons caused by aliovalent Hf and Zr doping can strongly modulate the $\kappa_L$ of NbFeSb. A similar phonon softening was reported in $(Fe_{1-x}Co_x)Si$ system[46–48]. In thermoelectrics, aliovalent doping is generally employed to optimize carrier concentration and electrical performance, with the further revelation of its significant role in suppressing thermal transport, the selection of a good dopant is thereby worth rethinking by synergistically considering its effect on both electrical and thermal transport properties.

Furthermore, the role of heavy elements (either serving as aliovalent dopants or isoelectronic substitutions), in the thermal transport of TE materials, deserves more attention. Experimental verifications of the atom rattling in the cage-like framework were first carried out in $CeFe_4Sb_{12}$[49] and $Ba_8Ga_{16}Ge_{30}$[38] by inelastic neutron scattering, and subsequent calculations have been performed to understand how the avoided-crossing behavior affects the phonon structure in a variety of filled skutterudites[50–52]. Avoided-crossing behaviors were also theoretically reported in other TE materials with atypical cage structures to explain their low $\kappa_L$, such as $CsAg_5Te_3$[53], $Ba_2AuBi$[54], and $AgBi_3S_5$[55]. Differently, for NbFeSb, which does not have a caged framework, it is the heavy aliovalent Hf and isoelectronic Ta that bring about the avoided-crossing behaviors of acoustic and optical branches. Acoustic phonons with large group velocities are depressed, resulting in reduced group velocity and suppressed $\kappa_L$. It is of great significance to find that heavy elements can exhibit such "guest atom-pseudo cage framework" behavior in the HH system, indicating the feasibility of selecting suitable extrinsic atoms to produce a similar effect to suppress $\kappa_L$ in compounds with a simple crystal structure other than the cage framework structure. These results will help elucidate the thermal transport mechanism of many good TE materials with suppressed $\kappa_L$, and also provide a means to suppress the $\kappa_L$ of thermal coating materials.

In summary, the effects of different aliovalent doping and isoelectronic alloying elements on phonon dispersion and phonon-phonon scattering of the heavy-band NbFeSb compounds are comprehensively investigated. Aliovalent dopants, Hf and Zr, introduce free charge carriers and enhance the screening, collapsing the LO-TO splitting and bringing the softening of high-energy optical phonons. The softening of optical phonons decelerates the optical phonons and thereby suppresses $\kappa_L$ effectively. The heavy elements, Hf and Ta, can bring about the avoided crossing of optical and acoustic phonon branches in the non-caged framework of the NbFeSb matrix. The depressed acoustic phonon branches with reduced group velocities at the energy range of 7-15 meV contribute to the suppressed $\kappa_L$. As a dopant with both aliovalent and heavy-element characteristics, Hf could induce both the softening of optical phonons and avoided crossing behavior of acoustic phonons in NbFeSb, resulting in the suppression of the group velocity in the full energy range and the largest drop of $\kappa_L$. Notably, a 65% reduction in the room-temperature $\kappa_L$ of NbFeSb by only 10% Hf doping, whereas, to achieve such a large drop, the content of isoelectronic Ta-alloying is 40%. Our work unravels the significant role of aliovalent dopants in phonon transport, in addition to their primary role in optimizing electrical performance, providing insights into the development of high-efficiency TE materials and thermal coating materials.

**Methods**

**Synthesis.** The ingots with nominal composition Nb$_{0.9}$X$_{0.1}$FeSb ($X$ = Nb, Ta, Hf, Zr) were prepared by levitation melting of the stoichiometric amount of Nb (foil, 99.8%), Zr (rod, 99.99%), Hf (slice, 99.99%), Ta (foil, 99.8%), Fe (block, 99.99%) and Sb (block, 99.999%) under an argon atmosphere for 3 minutes. Each ingot was remelted four times to ensure homogeneity and no raw material remained. The ingots were grounded manually into powders and mechanically milled (SPEX-8000D, PYNN Corporation) for 30 minutes under argon protection. The fine powders were compacted by SPS (LABOX-650F, Sinter Land Inc.) at 1123K under 65 MPa under vacuum for 10 minutes. The as-sintered samples were annealed at 1073K for 3 days. The chemical compositions of all sintered samples were checked by the electron probe microanalyzer (EPMA, JEOL JXA-8100) with a wavelength dispersive spectroscope (WDS).

**Measurements.** The electrical and thermal transport properties below 300K are characterized by the Physical Property Measurement System (PPMS, Quantum Design) with the electrical transport option and thermal transport options. The Seebeck coefficient and electrical conductivity from 300 to 1100K were measured on a commercial Linseis LSR-3 system with accuracy is ±5% and ±3%, respectively. The thermal conductivity $\kappa$ was calculated by using $\kappa = D\rho C_p$, where $\rho$ is the sample density estimated by the Archimedes method and $D$ is the thermal diffusivity measured by a laser flash method on Netzsch LFA457 instrument with a Pycoceram standard. The specific heat at constant pressure $C_p$ was calculated by using $C_p = C_{ph, H} + C_D$, where $C_{ph, H}$ and $C_D$ can be calculated by sound velocity, thermal expansion coefficient, and density. Normal and shear ultrasonic measurements were performed at room temperature using input from a Panametrics 5052 pulser/receiver with a filter at 0.03 MHz. The response was recorded via a Tektronix TDS5054B-NV digital oscilloscope. The thermal expansion coefficient was measured by Netzsch DIL 402 PC.

**Neutron powder diffraction.** The samples of about 3 grams were ground into powders in an agate mortar and used for NPD measurements. The measurements were taken on the High Resolution Powder Diffractometer for Thermal Neutrons, HRPT, at the Swiss Spallation Neutron Source (SINQ) of Paul Scherrer Institut (PSI), and on the Super High Resolution Powder Diffractometer, BL08 SuperHRPD, at the Material and Life Science Experimental Facility (MLF) of Japan Proton Accelerator Research Complex (J-PARC). Vanadium sample cans were used in the measurements. The NPD patterns were collected over the temperature range of ~10 K to ~300 K. Rietveld refinements were performed using the FullProf and Z-Rietveld softwares for the HRPT and SuperHRPD data, respectively.

**Inelastic neutron scattering.** The inelastic neutron scattering measurements were performed using the time-of-flight chopper spectrometer, MARI, at the ISIS Neutron and Muon Source in the UK. 5-10 grams of powder samples were wrapped in aluminum foil and then sealed in thin-walled cylindrical Al can, filled with low-pressure helium gas. The incident neutron energy, $E_i$ = 65 meV, was used with an energy resolution (full width at half maximum) of $\Delta E/E_i$ ~4.3% at the elastic line. The measurements were performed at 5 K, 100 K, 200 K, and 300 K with a closed-cycle He refrigerator. Correspondingly, the empty Al can was also measured in identical conditions at all temperatures. All of the spectra were normalized with respect to the scattering from a standard vanadium sample. The time-of-flight data were reduced with the MantidPlot. Then, the obtained powder-averaged dynamical structure factor $S(Q, E)$ was analyzed in the incoherent-scattering approximation after subtracting background, multi-phonon and multiple scattering as well as the elastic peak using the DAVE[56] and GetDOS programs[57]. Integration of the $S(Q, E)$ spectra over the range of momentum transfers $0.3 \leq Q \leq 11.5$ Å$^{-1}$ leads to neutron-weighted phonon DOSs for the samples[58].

**First-principles calculations.** The *ab initio* calculations were performed by using the projector augmented wave method, as implemented in the Vienna *ab initio* simulation package[59,60]. The generalized gradient approximation was used for the exchange-correlation functional[61], and a plane-wave energy cutoff of 450 eV was adopted. The phonon DOS and dispersions of NbFeSb and Nb$_{0.875}$X$_{0.125}$FeSb ($X$ = Ta, Hf, Zr) were calculated by using the frozen phonon method, as implemented in the Phonopy package[24]. The phonon scattering rates and lattice thermal conductivities were obtained via the shengBTE package, in which we considered the interactions between atoms to their fourth nearest neighbors[25]. A $2 \times 2 \times 2$ of the NbFeSb pritimive cell, with 24 atoms in total, is adopted to accomodate the 1/8 content of dopants in this work. For the structural optimization of the unit cell, the k-points were chosen as $3 \times 3 \times 3$, and the convergence accuracy of force was $10^{-5}$ eV/Å. For the supercells with displacements ($2 \times 2 \times 2$ of the 24-atom-cell, 192 atoms in total), which were used in the calculations of second- and third-order force constants, only Γ point was considered. The energy convergence criterion was $10^{-7}$ eV throughout the work. Besides, the electron-phonon interaction of primitive cell was performed by using Quantum ESPRESSO package[26] and EPW package[27] with norm-conserving pseudopotentials and Perdew-Burke-Ernzerhof exchange and correlation functional. A plane wave cutoff of 100 Ry is employed, the initial **k** and **q** meshes are both $6 \times 6 \times 6$, which are interpolated to $40 \times 40 \times 40$ meshes in order to calculate the electron-phonon interaction matrix. Due to the extremely time-consuming workload of the calculations for the electron-phonon interaction direct on the 24-atom-cell, in this work, we adopt an alternative procedure. The frequency-dependent electron-phonon (under the hole carrier concentration of $2.3 \times 10^{21}$ cm$^{-3}$, equivalent to the 1/8 aliovalent doping) and phonon-phonon scattering rates were calculated for the 3-atom primitive cell of NbFeSb, and thus the frequency-dependent reduction rate due to the introduction of ep could be obtained. The reduction rate was assumed to be the same for doped systems, and applied in the frequency-dependent thermal conductivity of the doped 24-atom-cell, as shown in Fig. 2c.

**Acknowledgments**

This work was supported by the National Science Fund for Distinguished Young Scholars (No. 51725102), the National Natural Science Foundation of China (Nos. 92163203, 52101275, and 52172216), and the Deutsche Forschungsgemeinschaft (DFG, German Research Foundation)—Projektnummer (392228380). J. Yang acknowledges the support of the Key Research Project of Zhejiang Lab (No. 2021PE0AC02). Experiments at the ISIS Neutron and Muon Source were supported by a beamtime allocation RB1820186 from the Science and Technology Facilities Council. Data is available here: https://doi.org/10.5286/ISIS.E.RB1820186. The authors thank Dr. Junjie Yu and Dr. Kaiyang Xia for the helpful discussions.

**Author contributions**

C.Fu and T.Z. designed the project. S.H. prepared the samples and carried out the high-temperature transport measurements. C.Fu and C.Felser performed low-temperature transport measurements. S.D. and J.Y. performed first-principles calculations. Q.R. and J.M. carried out the INS experiment and high-resolution neutron diffraction studies with input from M.L., D.S., P.M., S.T., and T.K.. S.H.

# Supplementary Information

**This Supplementary file includes:**

FIG. S1 to S10

**FIG. S1. Schematic illustration of optimal carrier concentrations at different DOS effective masses.**
**FIG. S2. Neutron powder diffraction patterns** for $Nb_{0.9}X_{0.1}FeSb$ (X = Nb, Ta, Hf, Zr) at 300K.
**FIG. S3. Electrical transport properties of** $Nb_{0.9}X_{0.1}FeSb$ ($X$ = Nb, Ta, Hf, Zr) **as a function of temperature.**
**FIG. S4. The calculated lattice thermal conductivity with the electron-phonon interaction.**
**FIG. S5. Experimental dynamical structure factor**.
**FIG. S6. Neutron-weighted phonon DOSs** for $Nb_{0.9}X_{0.1}FeSb$ (X = Nb, Ta, Hf, Zr).
**FIG. S7. First-principles calculation of the phonon dispersions** in the 2 × 2 × 2 supercell.
**FIG. S8. Calculated partial crystal orbital Hamilton population (pCOHP)** for $Nb_{0.9}X_{0.1}FeSb$ (X = Nb, Hf, Zr).
**FIG. S9. Calculated neutron-weighted partial phonon DOS** for $Nb_{0.875}X_{0.125}FeSb$ (X = Nb, Ta, Hf, Zr) in the 2 × 2 × 2 super cell.
**FIG. S10. Calculated partial phonon DOS for $X$ ($X$ = Ta, Hf, Zr) in $Nb_{0.875}X_{0.125}FeSb$ in full energy range.**

TABLE S1 to S5

**TABLE S1** Optimal carrier concentration and nominal doping concentration in experiments for typical TE materials.
**TABLE S2** Suppression of lattice thermal conductivity via aliovalent doping for typical TE materials.
**TABLE S3** Suppression of lattice thermal conductivity via isoelectronic alloying for typical HH materials.
**TABLE S4 The EPMA composition and relative density** of $Nb_{0.9}X_{0.1}FeSb$ ($X$ = Nb, Ta, Hf, Zr).
**TABLE S5 The calculated lattice thermal conductivity** of $Nb_{0.875}X_{0.125}FeSb$ ($X$ = Nb, Ta, Hf, Zr) with considering the electron-phonon scattering at 300K.

Note

**The enhancement of phonon-phonon scattering phase space**
**The effect of atomic mass on the avoided-crossing**

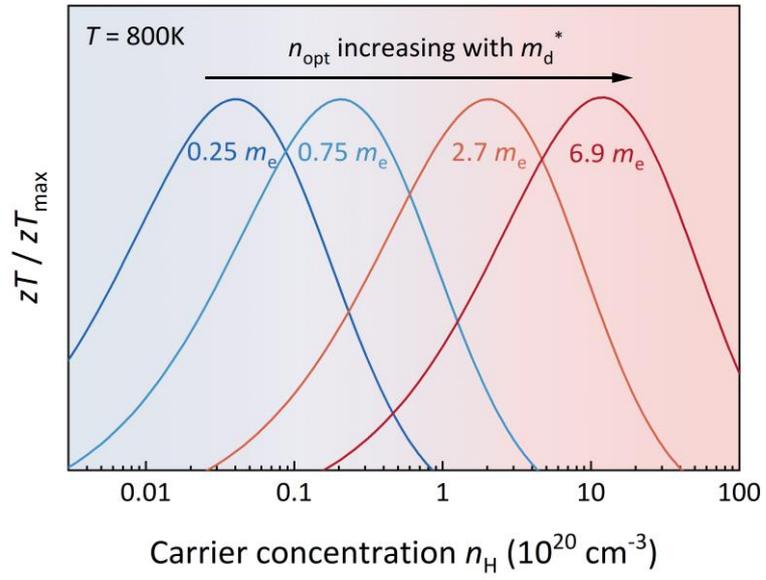

FIG. S1. Schematic illustration of optimal carrier concentrations at different DOS effective masses.

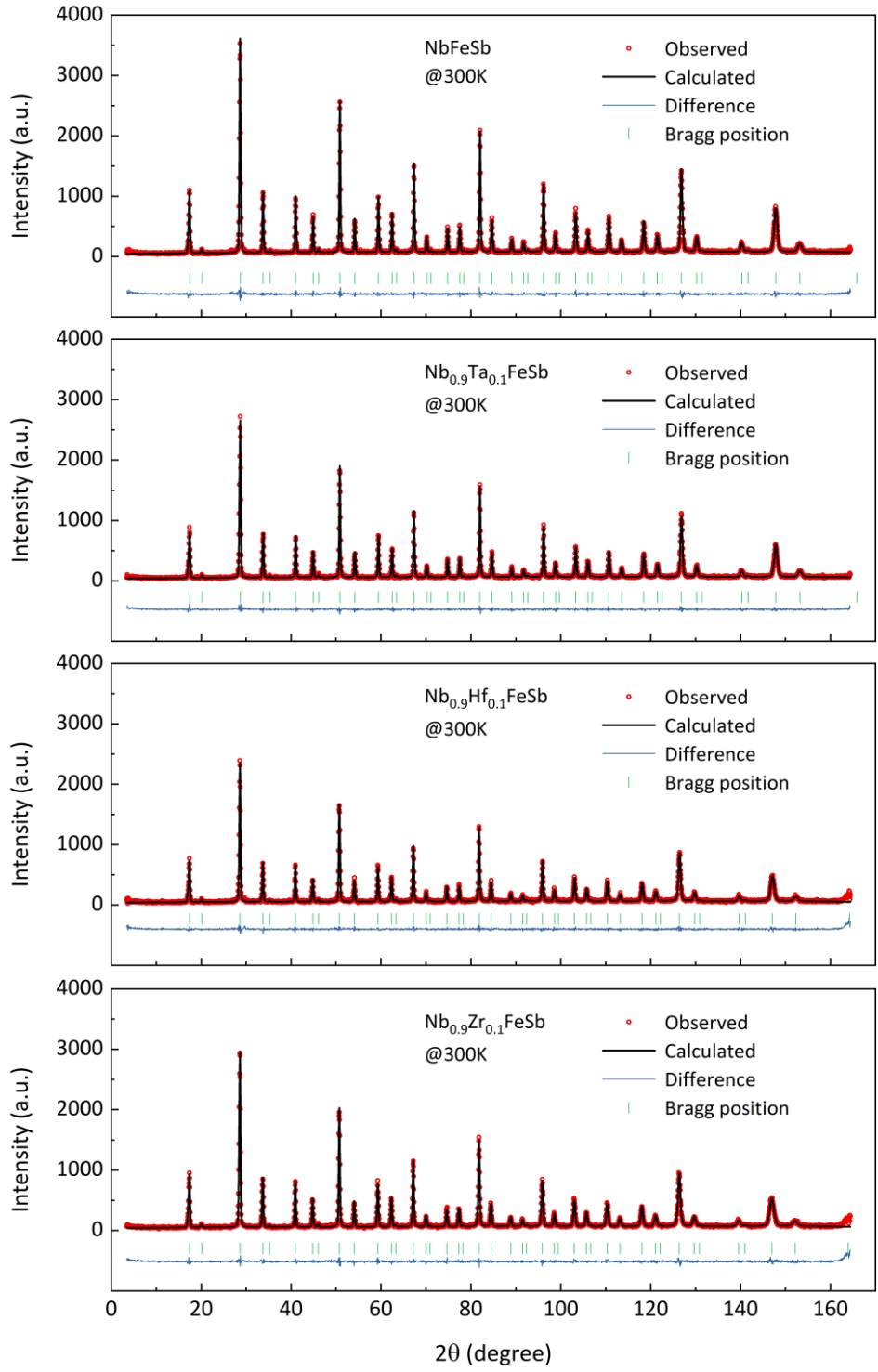

**FIG. S2. Neutron powder diffraction patterns** for $Nb_{0.9}X_{0.1}FeSb$ (X = Nb, Ta, Hf, Zr) at 300K.

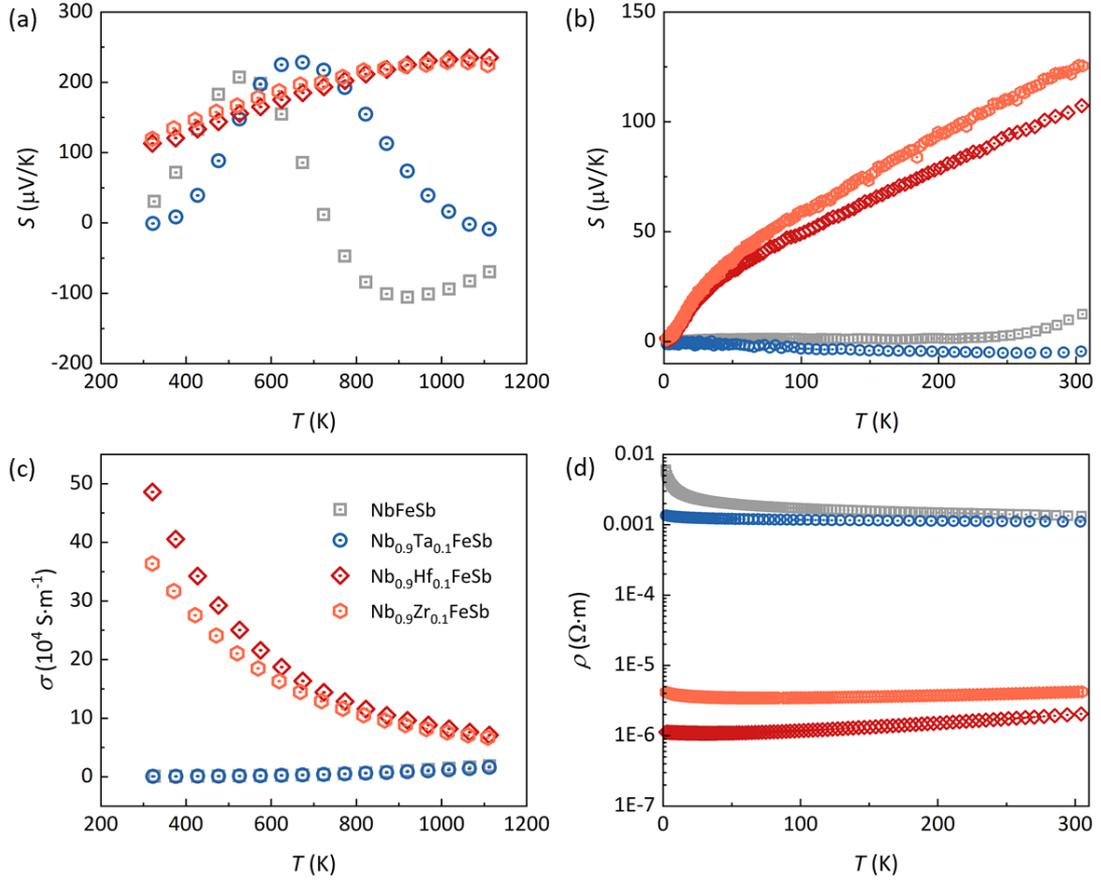

**FIG. S3. Electrical transport properties of** $Nb_{0.9}X_{0.1}FeSb$ ($X$ = Nb, Ta, Hf, Zr) **as a function of temperature.** Seebeck coefficient $S$ (a) above 300K and (b) below 300K, (c) electrical conductivity above 300K, and (d) electrical resistivity below 300K.

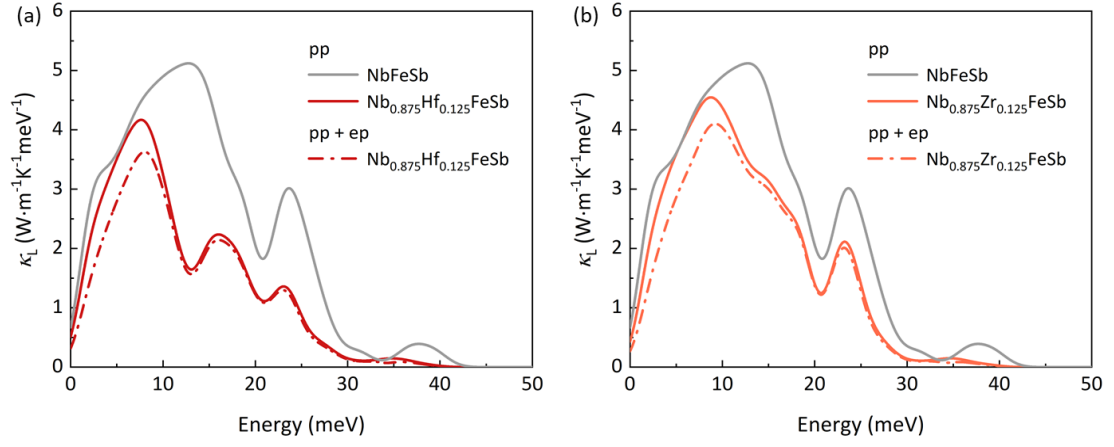

**FIG. S4. The calculated lattice thermal conductivity with the electron-phonon interaction** for (a) Nb$_{0.875}$Hf$_{0.125}$FeSb and (b) Nb$_{0.875}$Zr$_{0.125}$FeSb. The pp refers to the phonon-phonon scattering and ep refers to the electron-phonon scattering.

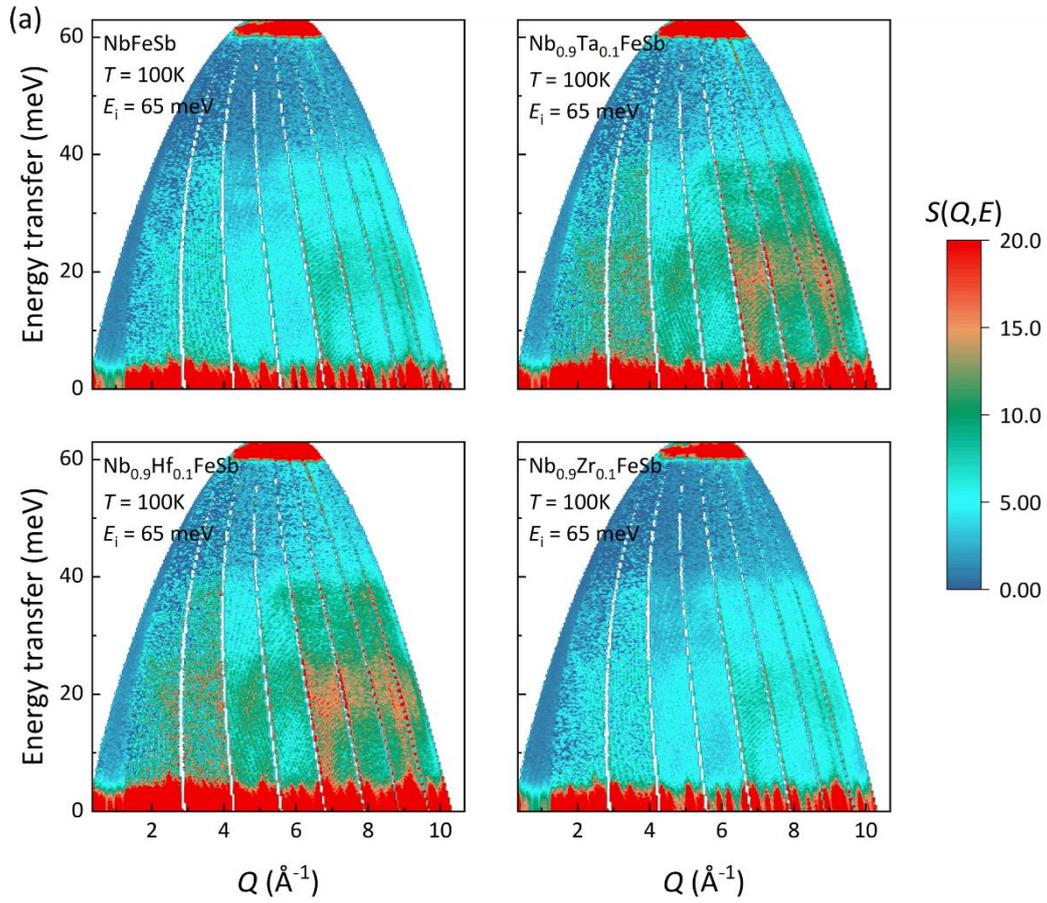

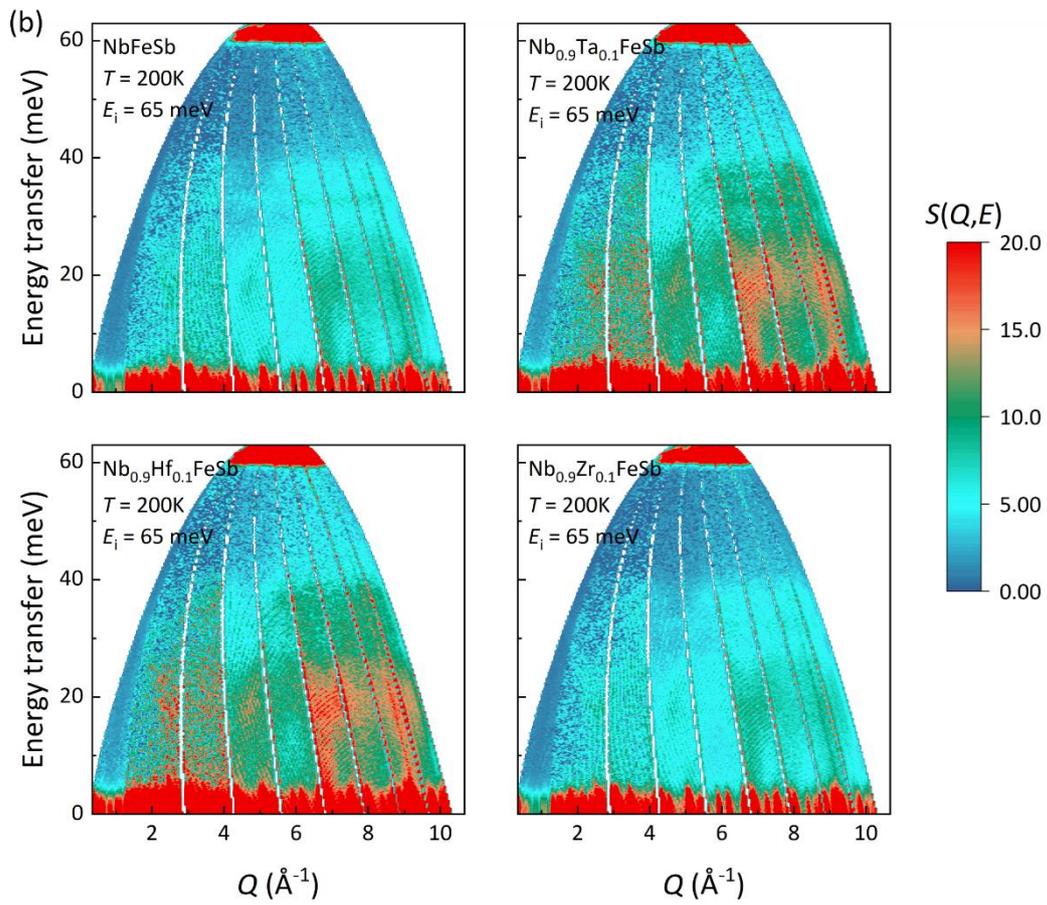

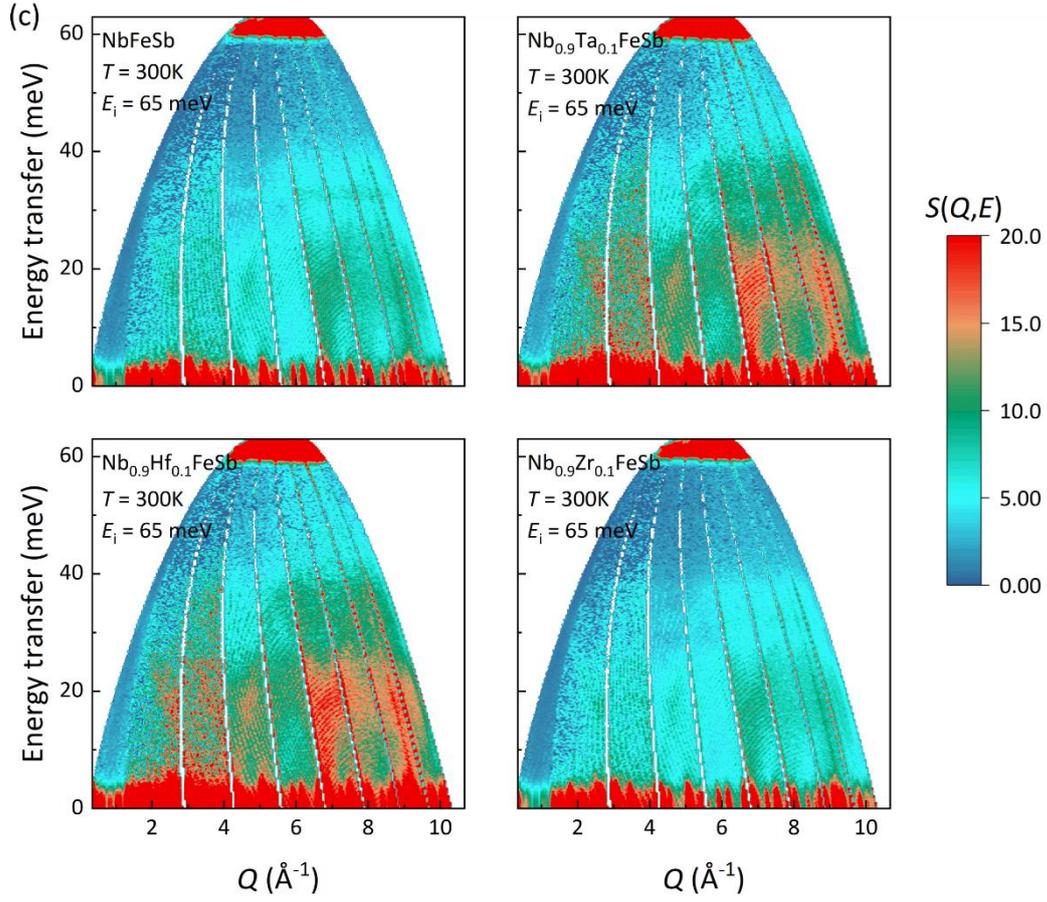

**FIG. S5. Experimental dynamical structure factor**, $S(\mathbf{Q}, E)$ for $Nb_{0.9}X_{0.1}FeSb$ ($X$ = Nb, Ta, Hf, Zr) polycrystalline samples at (a) 100 K, (b) 200K and (c) 300K with $E_i$ = 65 meV.

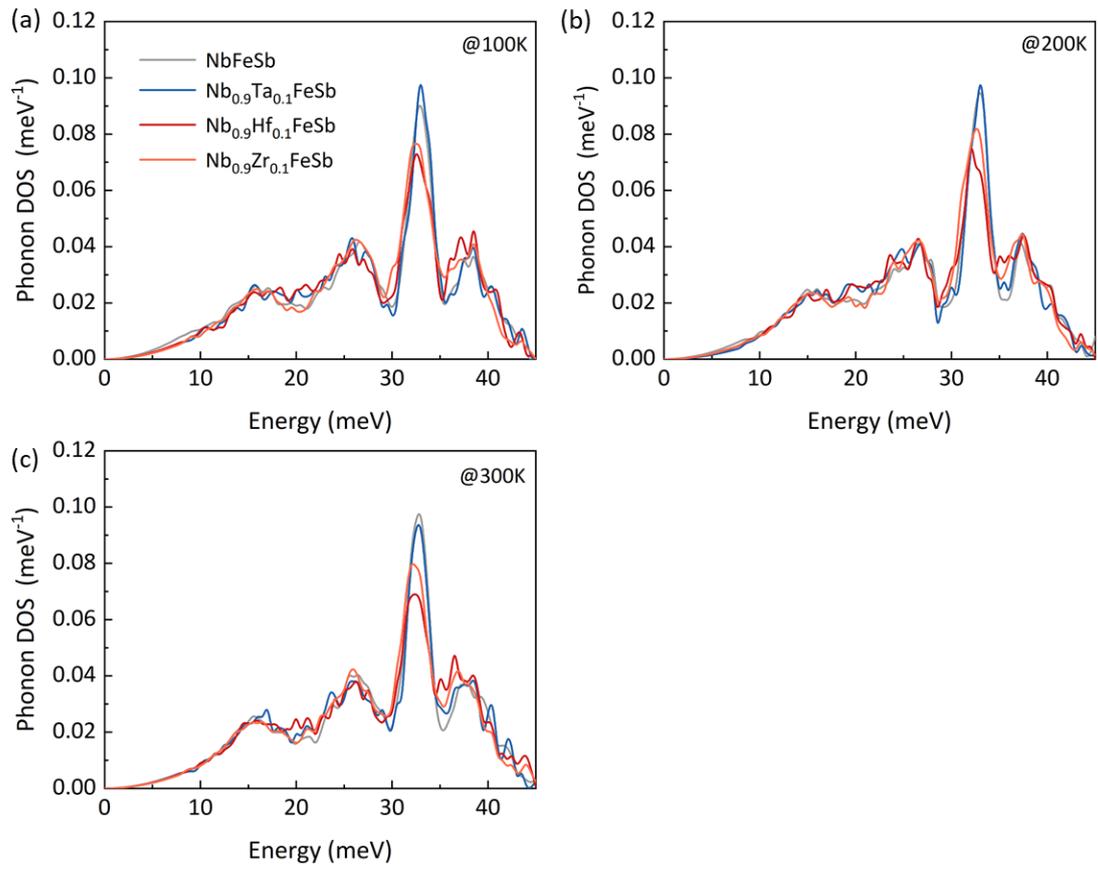

**FIG. S6. Neutron-weighted phonon DOSs** for Nb$_{0.9}$X$_{0.1}$FeSb (X = Nb, Ta, Hf, Zr) (a) at 100K, (b) at 200K, and (c) at 300K.

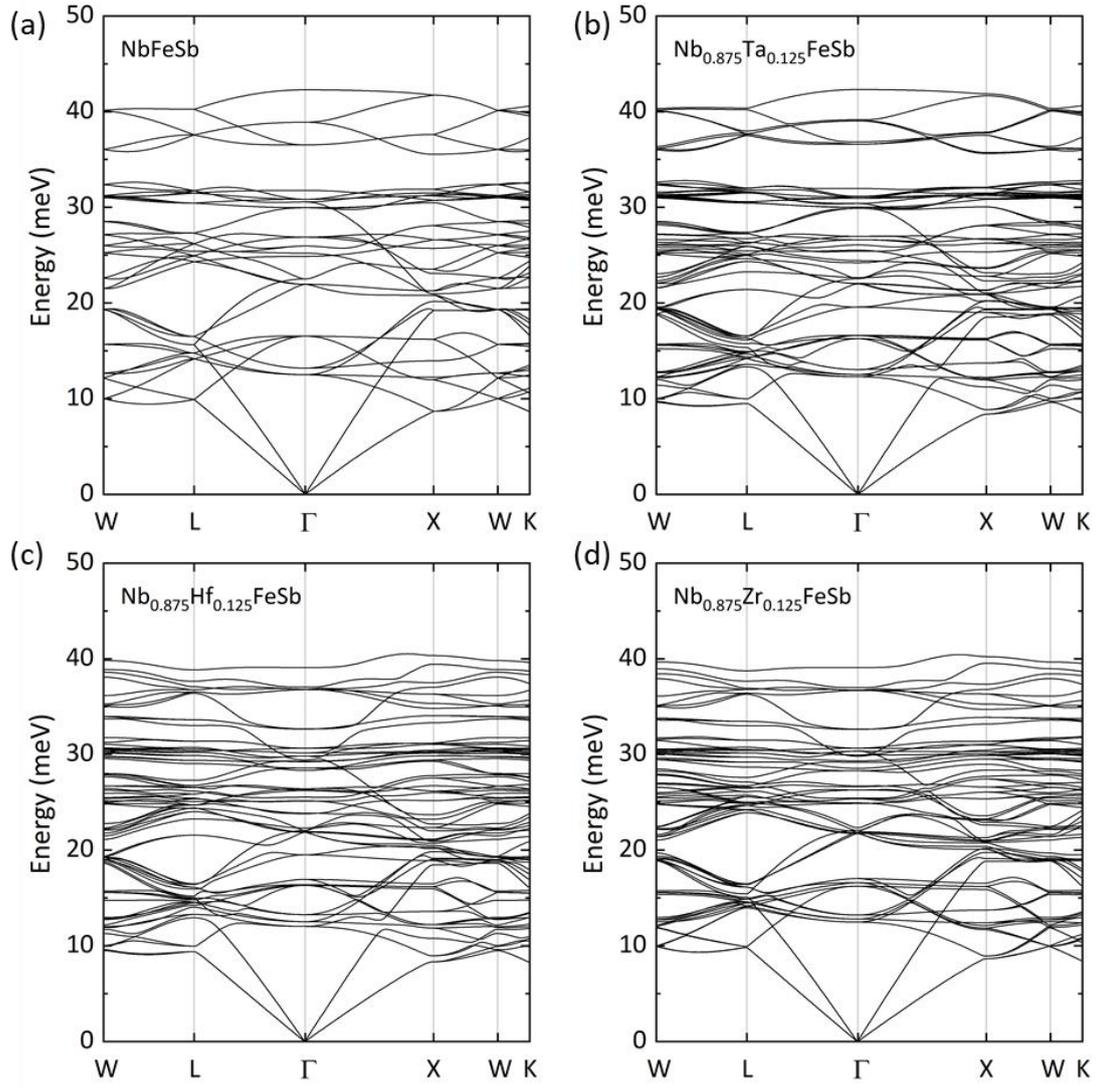

**FIG. S7. First-principles calculation of the phonon dispersions** in the supercell (a) NbFeSb, (b) $Nb_{0.875}Ta_{0.125}FeSb$, (c) $Nb_{0.875}Hf_{0.125}FeSb$, and (d) $Nb_{0.875}Zr_{0.125}FeSb$.

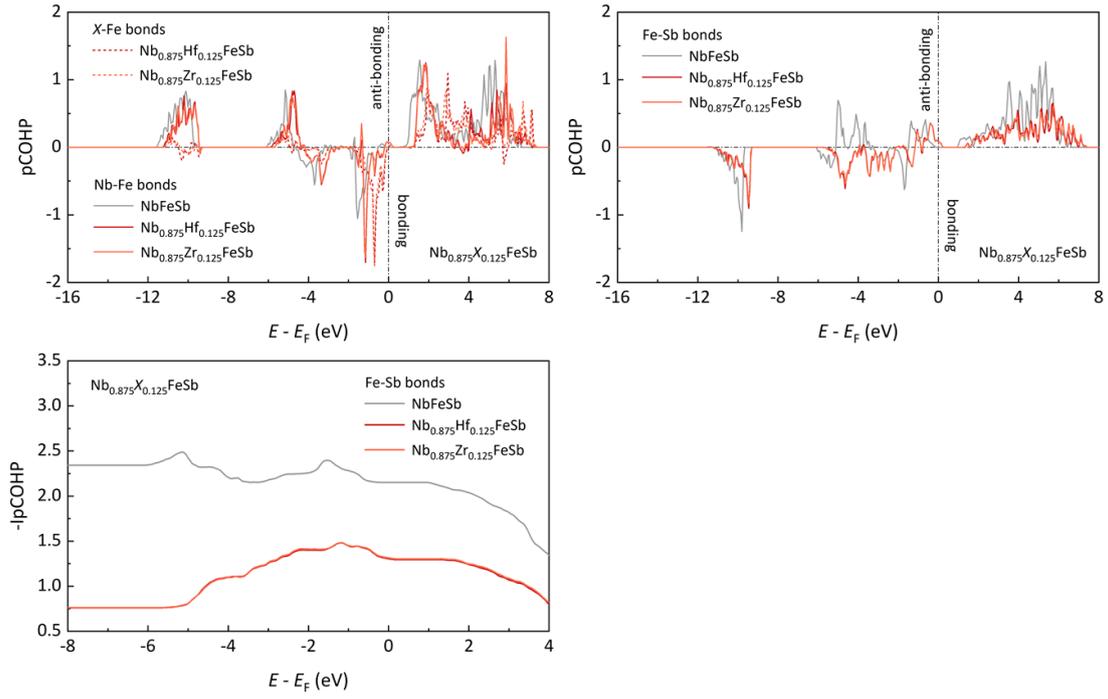

**FIG. S8.** Calculated partial crystal orbital Hamilton population (pCOHP) for $Nb_{0.9}X_{0.1}FeSb$ (X = Nb, Hf, Zr). pCOHP for the nearest neighbor (a) (Nb, Hf, Zr)-Fe bonds and (b) Fe-Sb bonds; (c) the negative integrated pCOHP (-IpCOHP) for the nearest neighbor Fe-Sb bonds.

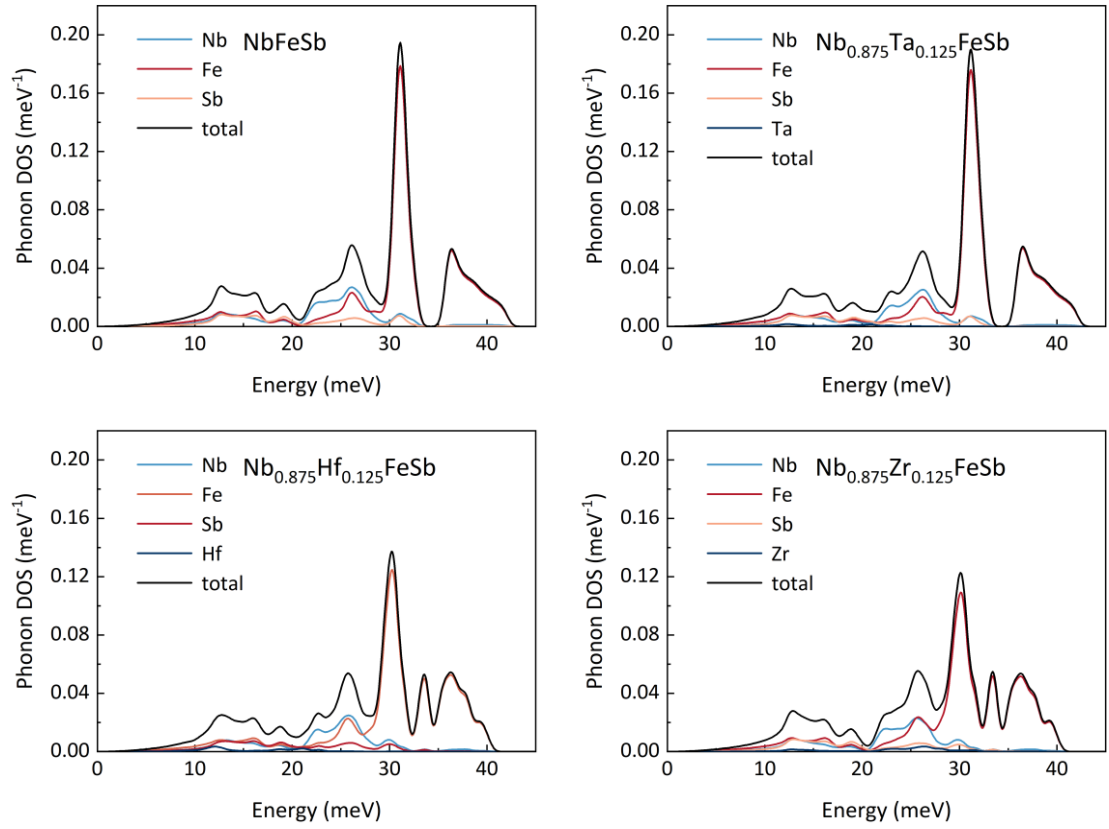

**FIG. S9. Calculated neutron-weighted partial phonon DOS** for $Nb_{0.875}X_{0.125}FeSb$ (X = Nb, Ta, Hf, Zr) in the super cell.

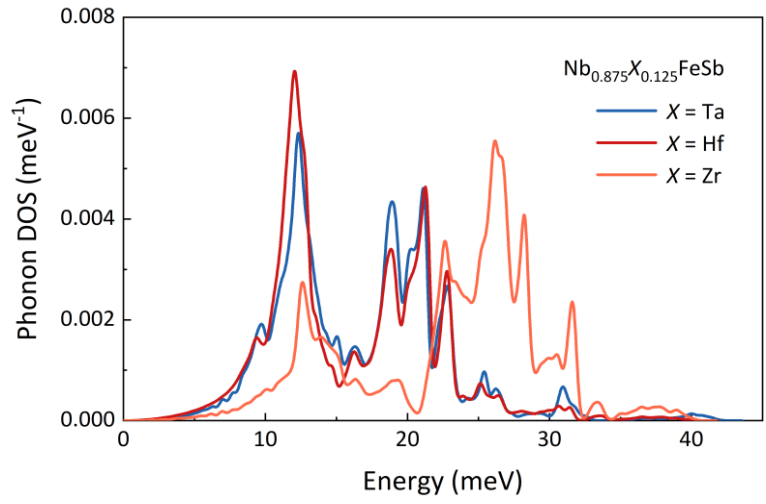

**FIG. S10.** Calculated partial phonon DOS for *X* (*X* = **Ta, Hf, Zr**) in **Nb$_{0.875}$X$_{0.125}$FeSb** in full energy range.

**TABLE S1** Optimal carrier concentration and nominal doping concentration in experiments for typical TE materials.

| Composition | $m_d^*$ ($m_e$) | $n_{opt}$ ($10^{20}$ cm$^{-3}$) | Nominal doping concentration | n/p type | Reference |
|---|---|---|---|---|---|
| PbTe$_{0.9988}$I$_{0.0012}$ | 0.25 | 0.17 | 0.12% | n | 1 |
| Pb$_{1.002}$Se$_{0.9982}$Br$_{0.0018}$ | 0.27 | 0.3 | 0.18% | n | 2 |
| PbS$_{0.9978}$Cl$_{0.0022}$ | 0.39 | 0.4 | 0.22% | n | 3 |
| Mg$_{2.99}$Ag$_{0.01}$Sb$_2$ | 0.7 | 0.4 | 0.33% | p | 4 |
| Mg$_3$(Sb$_{1.5}$Bi$_{0.5}$)$_{0.96}$Te$_{0.04}$ | 1.4 | 0.48 | 2% | n | 5 |
| BaGa$_{1.95}$Zn$_{0.05}$Sb$_2$ | 1.8 | 1 | 2.5% | p | 6 |
| Ba$_8$Ga$_{15.5}$Ge$_{30.5}$ | 1.86 | 4.3 | 1.67% | n | 7 |
| MgAgSb$_{0.99}$In$_{0.01}$ | 2 | 0.9 | 1% | P | 8 |
| Mg$_{2.16}$Si$_{0.45}$Sn$_{0.537}$Sb$_{0.013}$ | 2 | 1.8 | 1.3% | n | 9 |
| Zr$_{0.2}$Hf$_{0.8}$NiSn$_{0.985}$Sb$_{0.015}$ | 2.9 | 4 | 1.5% | n | 10 |
| Yb$_{0.3}$Co$_4$Sb$_{12}$ | 4.3 | 3.9 | 30% | n | 11 |
| Bi$_{0.94}$Pb$_{0.06}$CuSeO | 4.84 | 3.4 | 6% | p | 12 |
| Mo$_3$Sb$_{5.2}$Te$_{1.8}$ | 5.5 | 22 | 25.7% | p | 13 |
| (Zr$_{0.4}$Hf$_{0.6}$)$_{0.88}$Nb$_{0.12}$CoSb | 6.6 | 16 | 12% | n | 14 |
| Nb$_{0.88}$Hf$_{0.12}$FeSb | 6.9 | 20 | 12% | p | 15 |
| (V$_{0.6}$Nb$_{0.4}$)$_{0.8}$Ti$_{0.2}$FeSb | 10 | 15 | 20% | p | 16 |
| Zr$_{0.5}$Hf$_{0.5}$CoSb$_{0.8}$Sn$_{0.2}$ | 12.5 | 22 | 20% | p | 17 |

**TABLE S2** Suppression of lattice thermal conductivity via aliovalent doping for typical TE materials.

| Composition | Nominal doping concentration $x$ | $\kappa_L$ (Wm$^{-1}$K$^{-1}$) | Change ratio of $\kappa_L$ | Reference |
|---|---|---|---|---|
| (Mg$_{1-x}$Ag$_x$)$_3$Sb$_2$ | 0.0033 | 1.31 | 0.94 | 4 |
| | 0.0067 | 1.25 | 0.90 | |
| | 0.0083 | 1.12 | 0.81 | |
| PbS$_{1-x}$Cl$_x$ | 0.0013 | 2.30 | 0.93 | 3 |
| | 0.0022 | 2.11 | 0.85 | |
| Pb$_{1-x}$Na$_x$Se | 0.005 | 1.84 | 0.88 | 18 |
| | 0.007 | 1.75 | 0.84 | |
| | 0.015 | 1.59 | 0.76 | |
| Nb$_{1-x}$Ti$_x$FeSb | 0.08 | 10.77 | 0.61 | 19 |
| | 0.12 | 8.82 | 0.50 | |
| | 0.16 | 5.96 | 0.34 | |
| | 0.20 | 4.53 | 0.26 | |
| Nb$_{1-x}$Hf$_x$FeSb | 0.08 | 7.55 | 0.43 | 15 |
| | 0.10 | 5.22 | 0.29 | |
| | 0.12 | 4.87 | 0.27 | |
| | 0.14 | 3.18 | 0.18 | |
| ZrCoBi$_{1-x}$Sn$_x$ | 0.05 | 5.03 | 0.56 | 20 |
| | 0.10 | 3.75 | 0.42 | |
| | 0.15 | 3.03 | 0.34 | |
| | 0.20 | 2.56 | 0.28 | |
| (Ba,La,Yb)$_x$Co$_4$Sb$_{12}$ | 0.13 | 1.86 | 0.20 | 21 |
| | 0.17 | 1.37 | 0.14 | |
| | 0.22 | 1.32 | 0.14 | |
| | 0.25 | 1.21 | 0.13 | |

**TABLE S3** Suppression of lattice thermal conductivity via isoelectronic alloying for typical HH materials.

| Composition | Nominal alloying concentration $x$ | $\kappa_L$ (Wm$^{-1}$K$^{-1}$) | Change ratio of $\kappa_L$ | Reference |
|---|---|---|---|---|
| PbTe$_{1-x}$Se$_x$ | 0 | 2.00 | 1 | [22] |
| | 0.10 | 1.39 | 0.69 | |
| | 0.15 | 1.24 | 0.62 | |
| | 0.20 | 1.19 | 0.60 | |
| | 0.25 | 1.17 | 0.58 | |
| Mg$_2$Si$_{1-x}$Se$_x$ | 0 | 4.00 | 1 | [23] |
| | 0.20 | 2.17 | 0.54 | |
| | 0.30 | 2.03 | 0.51 | |
| | 0.40 | 1.91 | 0.48 | |
| | 0.50 | 1.88 | 0.47 | |
| Mg$_3$(Sb$_{1-x}$Bi$_x$)$_2$ | 0 | 1.59 | 1 | [24,25] |
| | 0.05 | 1.32 | 0.83 | |
| | 0.075 | 1.23 | 0.78 | |
| | 0.10 | 1.16 | 0.73 | |
| | 0.20 | 0.94 | 0.59 | |
| | 0.35 | 0.87 | 0.55 | |
| (Nb$_{1-x}$Ta$_x$)$_{0.8}$Ti$_{0.2}$FeSb | 0 | 4.75 | 1 | [26] |
| | 0.12 | 3.48 | 0.73 | |
| | 0.20 | 2.95 | 0.62 | |
| | 0.36 | 2.12 | 0.45 | |
| | 0.40 | 1.76 | 0.37 | |
| (Zr$_{1-x}$Hf$_x$)NiSn$_{0.985}$Sb$_{0.015}$ | 0 | 7.85 | 1 | [10] |
| | 0.20 | 5.41 | 0.69 | |
| | 0.40 | 4.77 | 0.61 | |
| | 0.50 | 4.72 | 0.60 | |
| (Zr$_{1-x}$Hf$_x$)CoSb$_{0.8}$Sn$_{0.2}$ | 0 | 5.99 | 1 | [17] |
| | 0.20 | 3.95 | 0.66 | |

| | | |
|---|---|---|
| 0.40 | 3.22 | 0.54 |
| 0.50 | 2.88 | 0.48 |

**TABLE S4 The EPMA composition and relative density** of $Nb_{0.9}X_{0.1}FeSb$ ($X$ = Nb, Ta, Hf, Zr).

| Nominal composition | EPMA composition | Relative density |
|---|---|---|
| NbFeSb | $Nb_{0.999}Fe_{1.001}Sb_{1.000}$ | 99.20% |
| $Nb_{0.9}Ta_{0.1}FeSb$ | $Nb_{0.900}Ta_{0.099}Fe_{1.000}Sb_{1.001}$ | 98.22% |
| $Nb_{0.9}Hf_{0.1}FeSb$ | $Nb_{0.901}Hf_{0.099}Fe_{1.000}Sb_{1.000}$ | 98.76% |
| $Nb_{0.9}Zr_{0.1}FeSb$ | $Nb_{0.883}Zr_{0.094}Fe_{1.019}Sb_{1.004}$ | 98.74% |

**TABLE S5 The calculated lattice thermal conductivity** of $Nb_{0.875}X_{0.125}FeSb$ ($X$ = Nb, Ta, Hf, Zr) with considering the electron-phonon scattering at 300K. (pp: phonon-phonon scattering; ep: electron-phonon scattering)

| Nominal composition | $\kappa_L$ (Wm$^{-1}$K$^{-1}$) only pp | $\kappa_L$ (Wm$^{-1}$K$^{-1}$) pp + ep |
|---|---|---|
| NbFeSb | 23.85 | / |
| $Nb_{0.875}Ta_{0.125}FeSb$ | 16.65 | / |
| $Nb_{0.875}Hf_{0.125}FeSb$ | 14.59 | 12.68 |
| $Nb_{0.875}Zr_{0.125}FeSb$ | 18.21 | 16.12 |

**Supplementary Note:**

**The enhancement of phonon-phonon scattering phase space**

Besides the softening and deceleration of phonons, the aliovalent doping and isoelectronic alloying modulate the scattering process by enlarging the scattering phase space. Fig. S11a showed the calculated scattering phase space in $Nb_{0.875}X_{0.125}FeSb$ compared with that in the pure NbFeSb. The enhancement over full energy range can be observed, particularly for the Hf-doped $Nb_{0.875}Hf_{0.125}FeSb$. The most commonly occurring phonon scattering process in a solid is the so-called three-phonon process, in which two phonons with vectors $\mathbf{k}_1$ and $\mathbf{k}_2$ collide and create a new phonon $\mathbf{k}_3$, or vice versa. Taking the absorption process as an example, the conservation of momentum and energy follows the expressions:

$$\mathbf{k}_1 + \mathbf{k}_2 = \mathbf{k}_3 + \mathbf{G} \quad , \tag{2}$$

$$\hbar\omega_{\mathbf{k}_1} + \hbar\omega_{\mathbf{k}_2} = \hbar\omega_{\mathbf{k}_3} \quad , \tag{3}$$

where $\mathbf{G}$ is a reciprocal lattice vector. $\mathbf{G} = 0$ refers to the normal (N) process with no direct generation of thermal resistance and $\mathbf{G} \neq 0$ refers to the Umklapp (U) process with thermal resistance. Determined by the energy conservation, new phonon $\mathbf{k}_3$ can stay at the higher energy and whether the U process can occur depends on whether a phonon state exists at the corresponding energy or not.

To further understand the U process in the current cases, the scattering phase spaces of emission and absorption processes were separated to differentiate the roles played by isoelectronic Ta and the aliovalent Hf and Zr doping, as shown in Fig. S11b and S11c. The enhancement of the three-phonon emission phase space in the energy range of 25-35 meV was observed for aliovalence-doped $Nb_{0.875}Hf_{0.125}FeSb$ and $Nb_{0.875}Zr_{0.125}FeSb$, but is not obvious for the isoelectronic Ta-alloyed $Nb_{0.875}Ta_{0.125}FeSb$, corresponding to the softening of high-energy optical phonons (Fig. 3b).

The reduction in energy and flattening of the high-energy optical phonon branches provides more energy-matched phonon states for the three-phonon U process involving acoustic-optical scattering, enlarging the scattering phase space for acoustic phonons and thus contributing to the suppression of $\kappa_L$. It is worth noting that the coupling between the low-lying transverse optical phonons and the longitudinal acoustic phonons was also theoretically and experimentally investigated in explaining the intrinsically low $\kappa_L$ of PbTe[27,28], in which the calculated $\kappa_L$ shows a six-times increase if acoustic-optical scattering processes were removed. In contrast, the aliovalent doping-induced phonon softening was rarely observed, particularly for the light-band TE materials.

For the enhancement of the emission phase space introduced by aliovalent doping, a characteristic phonon mode at 30.75 meV was chosen to illustrate which energy range of phonon pairs make differences in $Nb_{0.875}Zr_{0.125}FeSb$ compared to NbFeSb. Starting from energy conservation, we divide the participating phonon pairs in the emission process into two groups, a phonon in 0-10 meV combined with the other phonon in 20-30 meV (marked as $W^-_{(0-10)}$), and two phonons in 10-20 meV (marked as $W^-_{(10-20)}$). In this way, all results can be described, as the summed target energy close to 30.75 meV.

Table S6 showed the calculated emission phase space contributed by phonon pairs in different energy ranges. The phase space contributed by phonon pairs in 10-20 meV shows little difference between NbFeSb and $Nb_{0.875}Zr_{0.125}FeSb$, while the main difference raising from 0.535 $ps^4/rad^4$ in NbFeSb to 0.679 $ps^4/rad^4$ in $Nb_{0.875}Zr_{0.125}FeSb$ reflected in the phonon pairs consisting of one phonon in 0-10 meV and the other phonon in 20-30 meV. This result is consistent with the contrast image of the calculated phonon energy-dependent lattice thermal conductivity (Fig. 2c in the main text). The suppression in the lattice thermal conductivity of 0-10 meV is mainly reflected in $Nb_{0.875}Zr_{0.125}FeSb$ and $Nb_{0.875}Hf_{0.125}FeSb$, indicating that the aliovalent doping plays a role in the enhancement of the scattering of low-energy acoustic phonons with the large group velocities.

**TABLE S6** Constitution of the emission phase space of phonon modes in 30.75 meV.

|  | $W^-_{total}$ ($ps^4/rad^4$) | $W^-_{(0-10)}$ ($ps^4/rad^4$) | $W^-_{(10-20)}$ ($ps^4/rad^4$) |
| --- | --- | --- | --- |
| NbFeSb | 1.530 | 0.535 | 0.995 |
| $Nb_{0.875}Zr_{0.125}FeSb$ | 1.704 | 0.679 | 1.025 |
| **Value added** | **0.174** | **0.138** | **0.030** |

For low-energy acoustic phonons, the difference in the scattering phase space appears in the absorption process. In Fig. S11c, the enhancement of the absorption phase space was observed in $Nb_{0.875}Ta_{0.125}FeSb$ and $Nb_{0.875}Hf_{0.125}FeSb$, while it is not significant in $Nb_{0.875}Zr_{0.125}FeSb$. The large mass of Hf and Ta increases the phonon DOSs in the low energy range and introduces lower energy optical phonons, enhancing the scattering of the acoustic phonons with large sound velocities. Hf, as an aliovalent and heavy dopant, plays a dual role in enhancing the scattering phase space both in high- and low-energies, corresponding to the significant changes in the full energy range (Fig. S11a).

For the difference introduced by heavy elements, we selected a phonon mode in 5.77 meV with a large absorption phase space difference in NbFeSb and $Nb_{0.875}Hf_{0.125}FeSb$ to study the corresponding three-phonon pairs. We also divide the participating phonon pairs in the absorption process distinguished by the highest-energy phonon into two groups, phonon energy lower than 20 meV (marked as $W^+_{(<20)}$) and higher than 20 meV (marked as $W^+_{(>20)}$). The upper limit of the energy of acoustic phonons in NbFeSb, about 20 meV (Fig. 4a), was chosen as a scale to distinguish the final state energy of three-phonon scattering.

Table S7 showed the constitution of the absorption phase space of phonon modes in 5.77 meV. The phase space contributed by phonon pairs with the highest energy below 20 meV contained most of the absorption phase space increase between NbFeSb and $Nb_{0.875}Hf_{0.125}FeSb$, as 0.402 $ps^4/rad^4$ in the total 0.466 $ps^4/rad^4$, while the remaining part with final state energy higher than 20 meV only showed little increase. The enhancement of the absorption phase space by heavy elements mainly comes from the change of the phonon structure in the energy range of acoustic phonon.

**TABLE S7** Constitution of the absorption phase space of phonon modes in 5.77 meV.

|  | $W^+_{total}$ ($ps^4/rad^4$) | $W^+_{(<20)}$ ($ps^4/rad^4$) | $W^+_{(>20)}$ ($ps^4/rad^4$) |
|---|---|---|---|
| NbFeSb | 2.829 | 2.137 | 0.692 |
| $Nb_{0.875}Hf_{0.125}FeSb$ | 3.295 | 2.539 | 0.756 |
| **Value added** | **0.466** | **0.402** | **0.064** |

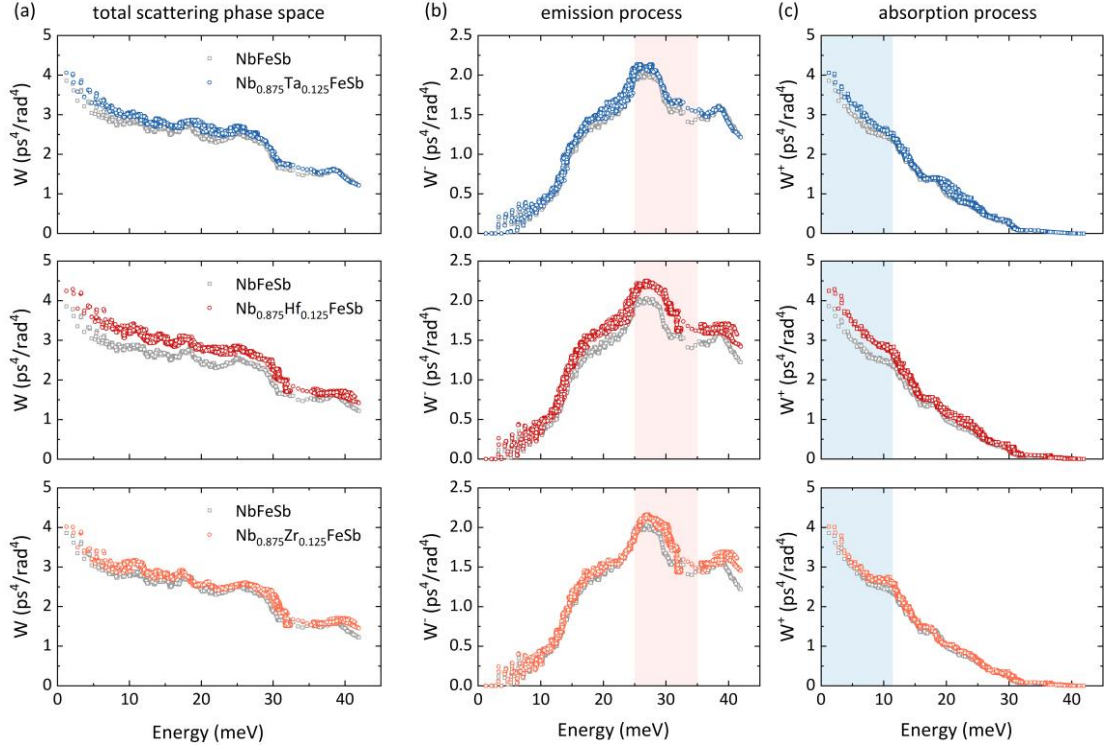

**Fig. S11. Energy dependence of phonon-phonon scattering phase space**. **a** total scattering phase space; **b** emission process (-); **c** absorption process (+). The Phonopy package[29] supports the description of specific three-phonon pairs. Characteristic phonon modes at 30.75 meV for the emission process and 5.77 meV for the absorption process were chosen to illustrate which energy range of phonon pairs corresponds to the specific enhancement.

**The effect of atomic mass on the avoided-crossing**

The separation degree of the avoided crossing increases with the mass of the heavy doping/alloying elements, which has been described using a linear chain model by Christensen et al.[30]. Defining a linear chain consisting of the cage wall with mass $M$ and the doping/alloying atoms with mass $m$. Each cage wall interacts with one another through the spring constant $K_1$ and the doping/alloying atom interacts only with the neighboring cage walls through a weaker spring constant $K_2$. The equations of motion can be written as:

$$M \frac{d^2}{dt^2} u_j(t) = K_1[u_{j-1}(t) + u_{j+1}(t) - 2u_j(t)] + K_2[v_{j-1}(t) + v_j(t) - 2u_j(t)] \quad , \quad \text{(S3)}$$

$$m \frac{d^2}{dt^2} v_j(t) = K_2[u_{j+1}(t) + u_j(t) - 2v_j(t)] \quad , \quad \text{(S4)}$$

In the case where $K_2 \ll K_1$ and $m \ll M$, defined the vibration frequencies as $\omega_1^2 = K_1/M$ and $\omega_2^2 = K_2/m$, the general solutions to the equation of motion can be taken the form:

$$u_j(t) = \alpha A \exp(iqjd - i\omega_q t) \quad , \quad \text{(S6)}$$

$$v_j(t) = A \exp(iq(j+1/2)d - i\omega_q t) \quad , \quad \text{(S4)}$$

By simplifying the mass difference by $\beta = m/M$, the equations for the two vibration frequencies can be obtained:

$$\omega_q^2 = 2\omega_1^2[1 - \cos(qd)] + 2\beta\omega_2^2[1 - \cos(qd/2)\alpha^{-1} - 2] \quad , \quad \text{(S7)}$$

$$\omega_q^2 = 2\omega_2^2[1 - \cos(qd/2)\alpha] \quad , \quad \text{(S8)}$$

By introducing $\gamma = \omega_1^2/\omega_2^2$, the final results can be found by inserting the solutions of $\alpha$ into either of the dispersion relations. The solutions of two values for $\alpha$ can be obtained:

$$\alpha = \frac{1 - \beta - \gamma[1 - \cos(qd)] \pm \sqrt{(1 - \beta - \gamma[1 - \cos(qd)])^2 + 4\beta \cos^2(qd/2)}}{2\cos(qd/2)} \quad , \text{(S9)}$$

Fig. S12(a) shows the separation degree increasing with $\beta$, meaning the increasing of $m$. For our NbFeSb system with dopant Hf, the $\beta = m_{Hf}/(4m_{Fe} + 6m_{Sb}) = 0.189$ is assumed by the simple basis of crystal structure, and the simulation dispersion relation is shown in Fig. S12(b). The shape of the avoided crossing is consistent with the phonon dispersion from first-principles calculations and the separation degree increases after we change the $m_{Hf}$ from 178.49 to a hypothetical value of 278, corresponding to the result marked with $\beta = 0.291$.

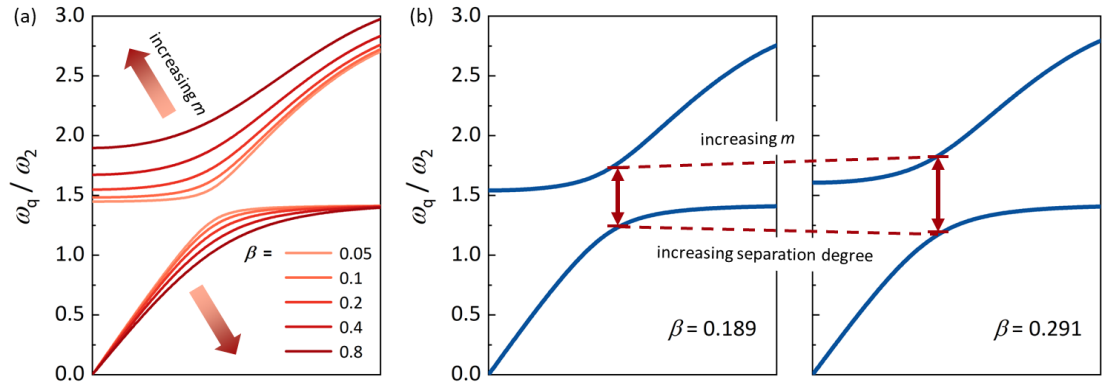

**FIG. S12. Schematic of dispersion relation calculated by linear chain model**. (a) The separation degree increases with the larger $\beta$; (b) comparison of the avoided crossing described by the real mass of Hf and a hypothetical mass 278. All these results were obtained by $\gamma = 2$ based on the original linear chain model.

To further verify the mass effect of the heavy doping/alloying elements, we changed the mass of Hf to 278 during the calculation manually in the Phonopy package and recalculated the phonon dispersion of $Nb_{0.875}Hf_{0.125}FeSb$. Fig. S13(a) shows the comparison in the L-Γ-X direction in the energy range of 0-25 meV. The acoustic branches show a further drop, which brings suppression for the group velocity of acoustic phonons. It is proved that Hf and Ta bring about the avoided crossing and the drop of group velocity due to their heavier mass, compared with Zr. The visualization of the phonon vibrational modes is further presented in Fig. S13(b). We chose two representative demonstrations of vibrational modes with avoided-crossing behaviors along Γ-L direction and Γ-X direction, exhibiting characteristics corresponding to longitudinal and transverse waves, respectively. The length of the arrow indicates the vibration amplitude of the corresponding atom, indicating the rattling-like behavior of the Hf atom in the pseudo cage framework of the NbFeSb matrix. (Visualization of the phonon vibrational modes through the TSS physics: http://henriquemiranda.github.io/phononwebsite)

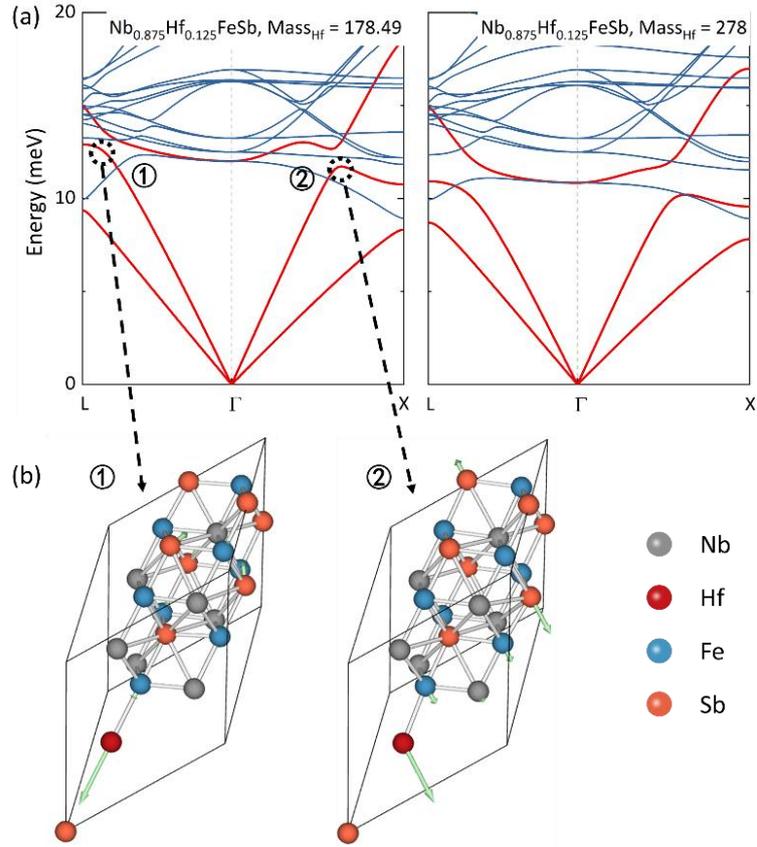

**FIG. S13. Calculated phonon dispersion and schematic diagram of atomic rattling-like modes for Nb$_{0.875}$Hf$_{0.125}$FeSb.** (a) Calculated phonon dispersion for Nb$_{0.875}$Hf$_{0.125}$FeSb with different settings of $m_{Hf}$; (b) schematic diagram of atomic rattling-like modes in Γ-L and Γ-X directions.